\documentclass[twocolumn]{aastex63} % for ApJ submission line numbers

\usepackage[colorinlistoftodos]{todonotes}
\usepackage[flushleft]{threeparttable}
\usepackage{graphicx}

\let\Oldtodo\todo
\renewcommand{\todo}[1]{\Oldtodo[inline]{#1}}

\hypersetup{linkcolor=cyan,citecolor=cyan,filecolor=cyan,urlcolor=cyan}

\usepackage{amsmath}
\usepackage{float}

\shorttitle{On the Formation of Planets in the Milky Way's Thick Disk}
\shortauthors{Hallatt \& Lee}

\graphicspath{{./}{figures/}}

\begin{document}

\title{On the Formation of Planets in the Milky Way's Thick Disk}

\correspondingauthor{Tim Hallatt}
\email{thallatt@mit.edu}

\author[0000-0003-4992-8427]{Tim Hallatt}
\affil{Department of Physics and Trottier Space Institute, McGill University, Montr\'eal, Qu\'ebec, H3A 2T8, Canada}
\affil{Institute for Research on Exoplanets (iREx), Montr\'eal, Qu\'ebec, Canada}

\author[0000-0002-1228-9820]{Eve J.~Lee}
\affil{Department of Physics and Trottier Space Institute, McGill University, Montr\'eal, Qu\'ebec, H3A 2T8, Canada}
\affil{Institute for Research on Exoplanets (iREx), Montr\'eal, Qu\'ebec, Canada}

\begin{abstract}

Exoplanet demographic surveys have revealed that close-in (${\lesssim}$1 au) small planets orbiting stars in the Milky Way's thick disk are ${\sim}50\%$ less abundant than those orbiting stars in the Galactic thin disk. One key difference between the two stellar populations is the time at which they emerged: thick disk stars are the likely product of cosmic noon (redshift $z {\sim}2$), an era characterized by high star formation rate, massive and dense molecular clouds, and strong supersonic turbulence. Solving for the background radiation field in these early star-forming regions, we demonstrate that protoplanetary disks at cosmic noon experienced radiation fields up to ${\sim}7$ orders of magnitude more intense than in solar neighborhood conditions. Coupling the radiation field to a one-dimensional protoplanetary disk evolution model, we find that external UV photoevaporation destroys protoplanetary disks in just ${\sim}$0.2--0.5 Myr, limiting the timescale over which planets can assemble. Disk temperatures exceed the sublimation temperatures of common volatile species for ${\gtrsim}$Myr timescales, predicting more spatial homogeneity in gas chemical composition. Our calculations imply that the deficit in planet occurrence around thick disk stars should be even more pronounced for giant planets, particularly those at wide orbital separations, predicting a higher rocky-to-giant planet ratio in the Galactic thick disk vs.~thin disk.

\end{abstract}

\section{Introduction} \label{sec:intro}

Approximately $\sim$30--50$\%$ of Sun-like stars are measured to host a super-Earth ($\sim$1--1.7$R_\oplus$) and/or a sub-Neptune ($\sim$1.7--4$R_\oplus$) inside $\sim$300 days \citep[e.g.,][]{howmarjoh10,batrowbry13,pethowmar13,donzhu13,frefrator13,rowbrymar14,burchrmul15,zhupetwu18}. With their occurrence rates far outweighing that of their larger counterparts, these small planets appear to be the most common type of planets among those that are discovered.
Around M dwarfs, the super-Earth/sub-Neptune population becomes even more ubiquitous with their occurrence rate enhanced by factors of $\sim$3 \citep[e.g.,][]{mulpasapa15,drecha15,gaimankra16,hsuforter20}. Given that these later-type stars are the most common type of stars in the Galaxy, planets smaller than $\sim$4$R_\oplus$ are likely the dominant outcome of Galactic planet formation.

Until recently, studies of the exoplanet population have been focused on those in the local 
solar neighborhood. With more precise measurements of stellar properties and exoplanet discoveries beyond the original {\it Kepler} field, there is evidence of changes in planet properties across different stellar populations across the Galaxy. For example,
combining stellar astrometric data from the \texttt{Gaia} spacecraft \citep{gaibroval18,gaibroval21}, radial velocities from \texttt{LAMOST} \citep{xiatinrix19}, spectroscopic data from the \texttt{GALAH} \citep{budaspduo18} and \texttt{APOGEE} \citep{majschfri17} surveys, and the planet sample from \texttt{Kepler} DR25 \citep{thocouhof18}, \citet{baszuc22} found a reduction by $\sim$50$\%$ in the occurrence rate of 0.5--10$R_\oplus$ planets within $\sim$300 days around FGK stars in the Milky Way's thick disk as compared to thin disk stars. 
\cite{zinharchr23} updated the stellar catalogs of \cite{harzinchr20} and \cite{berhubvan20} using stellar values from \texttt{LAMOST} Data Release 8 \cite[][]{wanhuayua22}, APOGEE Data Release 17 \cite[][]{abdaccaer22}, the California-Kepler Survey \cite[][]{pethowmar17}, and new K2 targets \cite[][]{petcroisa18}. Combined with the \texttt{Kepler} Data Release 25 planet sample (updated by \cite{berhubgai20} and \cite{zinharchri21}), \cite{zinharchr23} observed a 50$\pm$8$\%$ and 56$\pm$7$\%$ decrease in the super-Earth and sub-Neptune occurrence rate, respectively, within $\sim$40 days' orbital periods around thick disk stars. 

What may be the cause of the reduced population of the close-in super-Earths/sub-Neptunes around thick disk stars? Stars in the thin disk are $\sim$3 Gyr old, exhibit solar metallicity on average, and are kinematically cool. Thick disk stars on the other hand are $\sim$10 Gyr old, kinematically hot, have lower metallicities, and exhibit higher abundances of $\alpha$ elements relative to iron at a given metallicity ([Fe/H]$\sim$-0.5, [$\alpha$/Fe]$\sim$+0.3 \cite[][]{redlamall06}; a result of their birth in the presence of type II supernovae, but before Fe-synthesizing type Ia supernovae). \citet{zinharchr23} ruled out the low [Fe/H] being the dominant reason for the reduction in planet occurrence, as they found the dependence of occurrence on [Fe/H] only accounts for 14$\pm$3$\%$ and 19$\pm$2$\%$ occurrence reductions for super-Earths and sub-Neptunes, respectively. What little correlation exists between stellar [Fe/H] and sub-Neptune or super-Earth occurrence may arise from increasing close stellar binary occurrence towards lower [Fe/H], which can inhibit planet formation via dynamical effects \citep{kutwuqia21}. Figure B1 of \cite{mazbadmoe20} however shows that the close (periods ${\leq}10^{4}$ days) binary fraction for stars with thick disk compositions [Fe/H]${\sim}{-}0.4$, [$\alpha$/Fe]${\sim}{+}0.3$ ([$\alpha$/H]${\sim}{-}0.1$ in their plot) is similar to that of Solar-type, thin disk stars (${\sim}20\%$; see their Figure 10). Similar close binary fractions in thin and thick disks also points against stellar binarity as the origin for the deficit of planets around thick disk stars relative to thin disk.

Dynamical ejection and disruption via frequent stellar fly-bys are also possible as suggested by both \citet{baszuc22} and \citet{zinharchr23}, although as the latter authors caution, significant dynamical upheaval would be required to disrupt inner orbits ($\lesssim$40 days) which are deep in the potential well of the central star. Furthermore, as demonstrated by \citet{snahawdim14,snahaydim15}, stars in the thick disk likely originate from ``cosmic noon'' at redshift $z {\sim}2$ when the star formation rate was at its peak in cosmic history \citep{maddic14,forwuy20}. Dense stellar clusters born during this era are expected to be short-lived due to tidal perturbations by dense gas clumps \cite[e.g.][]{kru15} so it would be difficult to sustain a high enough rate of stellar fly-bys over $\sim$1--10 Gyr timescale to remove inner-orbit planets.

Instead, we consider a third possibility. The time available for planet formation is limited by the dissipation timescale of the protoplanetary disk. While purely rocky planets could assemble in completely gas-free environment, the onset of the initial protocore assembly is facilitated by dust-gas interaction including aerodynamic radial drift \cite[e.g.][]{wei77}, drag instabilities that locally concentrate dust \cite[e.g.][]{yougoo05,squhop18}, and pebble accretion \cite[e.g.][]{ormkla10,lamjoh12}. Furthermore, sub-Neptunes require $\sim$1--10 \% by mass H/He envelope \cite[e.g.][]{lopfor14,wollop15} which must originate from the primordial nebula. In radiatively extreme environments such as the Galactic Center, externally-driven photoevaporation has been shown to be effective in truncating disk lifetimes well within $\sim$1 Myr \citep[e.g.,][]{winkruche20,owelin23}. 

Given that thick disk stars likely originate from the era of maximal star formation activity, they are expected to have been subjected to similarly extreme radiation environments. We will investigate the time evolution of circumstellar disks in nebular conditions at redshift $\sim$2, characterized by strong supersonic turbulence \citep[see e.g.,][their Figure 2]{girfisbol21}, high cloud surface density and mass \citep[see e.g.,][their Figure 3 and suppementary Figure 6, respectively]{desriccom19},
and enhanced star formation efficiency \citep[see e.g.,][their Figures 2 and 4]{gruhopfau18}.

The paper is organized as follows. Section \ref{sec:model} outlines the model setup for calculating the radiation environment in the primordial thick disk and its feedback on the evolution of protoplanetary disks around member stars. Results of our calculations are presented in Section \ref{sec:results}. Section \ref{sec:discussion} explores the effect of varying level of extinction by dust and turbulent viscosity on our calculation, discusses the implication on the formation of gas giants, and provides several observational predictions.
We conclude in Section \ref{sec:conclusion}. 

\section{Stellar Cluster \texorpdfstring{$\&$}{and} Protoplanetary Disk Model}\label{sec:model}

Our goal is to compute the radiation field impinging on a random star embedded in a ``clustered" young stellar population inside a star-forming giant molecular cloud (GMC). The properties of the galactic disk, GMC, stellar dynamics, stellar position correlation function, interstellar extinction, and therefore the radiation field, are assumed to follow from the density and turbulent statistics of the interstellar medium (ISM). We scale the ISM properties backwards in time from the present-day (when thin disk stars are born) to the time at which thick disk stars emerged ($\sim$10 Gyr ago) to estimate the extent to which the star and planet-forming environment may differ between the two stellar populations. We outline below the details of our model setup and summarize our parameter choices in Table \ref{param_table}.

\subsection{Stellar Clustering and Radiation}
\label{ssec:clustering}

Stars are born ``clustered", in a \textit{statistical} sense (not in the sense that clusters are necessarily gravitationally bound). The degree of spatial clustering is described by the autocorrelation function $\xi(r)$ \citep[e.g.][]{hop13,leehop20},

\begin{equation}\label{equation:correlation_function}
    1+\xi(r)=\frac{\big<N(r)\big>}{\big<n\big>dV}
\end{equation}

\noindent where $N(r)$ is the observed number of stars in a differential volume element $dV$ at distance $r$ from a random subject star ($r$ is \textit{not} the distance from the cloud center), $\big<n\big>$ is the mean number density of stars, and $\big<\hdots\big>$ denotes ensemble averaging. The correlation function therefore describes the excess probability of finding a star at distance $r$ from another star. Observations \citep[e.g.,][]{krahil08} and theories of ISM turbulence statistics \citep[e.g.,][]{hop13} both find that the correlation function of stars at birth scales approximately as a power law according to $1+\xi(r)\propto r^{-2}$ down to scales $\gtrsim 10^{-2}$ pc \citep[see, e.g.,][their Figure 1 and references therein]{leehop20} 
below which the correlation function steepens likely due to binary formation via disk fragmentation \citep[e.g.,][]{gushopkru17,gushopgru18}. We therefore take $1+\xi(r)\propto r^{-2}$ as an initial condition throughout the paper.

Following \cite{leehop20}, we compute the average unattenuated flux impinging on a random star as
\begin{equation}
    \big<F_\star\big> = \frac{L_{\rm cl}}{4\pi r_{\rm cl}^2}\frac{M_{\star,\rm cl}}{30\,M_\odot} = \frac{1}{4}\big<L_\star/m_\star\big>\epsilon_\star\Sigma_{\rm cl}\frac{M_{\star,\rm cl}}{30\,M_\odot}
    \label{eq:Fstar}
\end{equation}
where $L_{\rm cl}/4\pi r_{\rm cl}^2$ is the mean luminosity per unit area averaged over the entire volume of the cloud, $r_{\rm cl}$ is the radius of the cloud, $\big<L_\star/m_\star\big>$ is the specific luminosity averaged over the stellar initial mass function (IMF), $\epsilon_\star \equiv M_{\star,\rm cl}/M_{\rm cl}$ is the star formation efficiency, $M_{\star, \rm cl}$ is the total stellar mass in the GMC, $M_{\rm cl}$ is the total mass (star and gas) in the GMC, and $\Sigma_{\rm cl} \equiv M_{\rm cl}/\pi r_{\rm cl}^2$ is the GMC mass surface density. We adopt the IMF of \citet{kro02} and \citet{kroweipfl13}:
\begin{align}\label{equation:IMF}
\begin{split}
    p(m_{\star})&\propto m^{-1.3}, \ 0.07\leq m_{\star}<0.5\\
                &\propto m^{-2.3}, \ 0.5\leq m_{\star}\leq 150.
\end{split}
\end{align}
The 30$M_\odot$ in the denominator of equation \ref{eq:Fstar} accounts for the sampling effect in the averaging, i.e., $r$ needs to be large enough to include enough stellar mass $\sim$30$M_\odot$ so that the IMF can be well-sampled.

As can be inferred from equation \ref{eq:Fstar}, the key consequence of clustering is that the incident flux on a given star enhances from a spatial average ($L_{\rm cl}/4\pi r_{\rm cl}^2$) by a factor $M_{\star,\rm cl}/30 M_\odot \propto \epsilon_\star M_{\rm cl}$. 
A typical star born in the massive ($\sim$10$^{7-9} M_{\odot}$), starburst GMCs which 
%may be characteristic of 
likely reflect the conditions of
the primordial thick disk \cite[e.g.][]{cladebnid19} will thus experience $\sim$1-3 orders of magnitude higher radiative flux than stars born in present-day Milky Way star-forming GMCs ($\sim$10$^{6} M_{\odot}$; \cite{leemivmur16,mivmurlee17}).

\subsection{Galactic Disk \texorpdfstring{$\&$}{and} GMC Properties}
\label{ssec:gal-disk-gmc}

We compute the expected cloud masses $M_{\rm cl}$ in the primordial thick disk using the Jeans mass under gravitational fragmentation of the galactic disk \cite[][]{kimost01,bintre08}:

\begin{equation}\label{equation:jeans}
\begin{split}
    M_{\rm J}&=\frac{\sigma_{\rm gal}^{4}}{G^{2}\Sigma_{\rm gal}} \\
             &{\sim}10^{8} \bigg(\frac{\sigma_{\rm gal}}{45 \ \mathrm{km}  \ \mathrm{s}^{-1}}\bigg)^{4}\bigg(\frac{1700 \ M_{\odot} \ \mathrm{pc}^{-2}}{\Sigma_{\rm gal}}\bigg) \ M_{\odot},
\end{split}
\end{equation}

\noindent with $\sigma_{\rm gal}$ the one-dimensional ISM gas velocity dispersion, and $\Sigma_{\rm gal}$ the ISM gas surface density.

For a galactic disk at redshift $\sim$2, we take $\Sigma_{\rm gal} = 1700 M_{\odot}$ pc$^{-2}$ (\cite{tacnergen13}; their Figure 7, left panel), and vary $\log_{10}{\sigma_{\rm gal}}\in  [1.40, \ 1.66, \ 1.83]$ km s$^{-1}$ \citep[e.g., see][their Figure 2, right panel]{girfisbol21}, to obtain $M_{\rm cl} = M_{\rm J} \in [10^{7},\ 10^{8}, \ 6{\times}10^{8}] \ M_{\odot}$. Here, the values of $\sigma_{\rm gal}$ are taken by visual inspection of Figure 2 of \citet{girfisbol21} at redshift 1, 2, and 3, using the approximate midpoint of the TNG50 simulation result from \citet{pilnelspr19} since they agree the best with the measurements reported by \citet{girfisbol21}. The available data only extend to redshift $\sim$2.47, so we use a least-squares fit to obtain an extrapolated value at redshift 3.

We now derive the velocity dispersion and the volumetric density of gas {\it within an  individual GMC}, denoted $\sigma_{\rm cl}$, and $\rho_{\rm cl}$, respectively.
Given the definition of virial parameter \citep{bermck92} $\alpha_{\rm vir} = 5\sigma_{\rm cl}^2 r_{\rm cl}/GM_{\rm cl}$, $M_{\rm cl} = M_{\rm J}$, and $\Sigma_{\rm cl} \equiv M_{\rm cl}/\pi r_{\rm cl}^2$ (the surface density of GMC), we can write
\begin{equation}\label{equation:sigcl_siggal}
    \sigma_{\rm cl} = (\pi \phi_\Sigma)^{1/4}\left(\frac{\alpha_{\rm vir}}{5}\right)^{1/2}\sigma_{\rm gal}
\end{equation}
where $\phi_\Sigma \equiv \Sigma_{\rm cl}/\Sigma_{\rm gal}$.

Under the virial theorem, the mean gas pressure inside the GMC can be written as
\begin{equation}
    P_{\rm cl} = \frac{3\pi}{20}\alpha_{\rm vir}G\Sigma_{\rm cl}^2
    \label{eq:Pcl}
\end{equation}
which, for a bound cloud $\alpha_{\rm vir}=1$, is simply a statement of the kinetic energy balancing self-gravity of spherical cloud \citep[e.g.,][]{mcktan03}. Since the gas flows are highly supersonic, gas kinetic energy is dominated by turbulence so $P_{\rm cl} = \rho_{\rm cl}\sigma_{\rm cl}^2$ and we arrive at
\begin{equation}
    \rho_{\rm cl} = \frac{3}{4}\pi^{1/2}\phi_\Sigma^{3/2}\frac{G\Sigma_{\rm gal}^2}{\sigma_{\rm gal}^2}.
    \label{equation:bulk_density}
\end{equation}

For both $\sigma_{\rm cl}$ and $\rho_{\rm cl}$, the remaining unknown parameter is $\phi_\Sigma$. 
For high redshift $\sim$2, obtaining high spatial resolution CO data is challenging. We take the median of the clump measurements available in limited cases for $z=1$--3 from \citet[][]{swipapcox11,swidyenig15,desriccom19,desriccom23} and find $\Sigma_{\rm cl} {\sim}3400\,M_\odot\,{\rm pc}^{-2}$, which leads to $\phi_\Sigma = 2$. With $\alpha_{\rm vir}=1$, 
$\sigma_{\rm cl} \simeq 0.71\sigma_{\rm gal}$ for the high redshift environment. As a result, our $\sigma_{\rm cl} \sim$17.8--47.9${\rm km\,s^{-1}}$ for $z$=1--3, which is in ballpark agreement with the measured CO linewidth of identified clumps at $z{\sim}1$, $\sim$15.5--21${\rm km\,s^{-1}}$ by \cite{desriccom19,desriccom23} and at $z{\sim}2$--3, $\sim$31--63${\rm km\,s^{-1}}$ by \cite{swipapcox11,swidyenig15}.

For present-day GMCs in the solar neighborhood, we take $\Sigma_{\rm cl}{\sim}60 M_{\odot} \rm pc^{-2}$ and $M_{\rm cl}{\sim}10^{5} M_{\odot}$ \citep[see the top left panel of Figure 7 in][and their Table 2]{mivmurlee17}, which is similar to the mass of the Orion clouds \citep[][]{madmormos86}. We compute $\sigma_{\rm cl}=(\alpha_{\rm vir} G M_{\rm cl}/5 r_{\rm cl})^{1/2} \simeq 1.9$ km s$^{-1}$, with $r_{\rm cl} = (M_{\rm cl}/\pi\Sigma_{\rm cl})^{1/2} {\sim}23$ pc, for solar neighborhood GMCs.

With the estimate of $\Sigma_{\rm cl}$, we can now choose a reasonable range of star formation efficiency $\epsilon_\star$. For solar neighborhood, we take the Milky Way average 2\%. For our thick disk progenitor condition, we adopt $\epsilon_\star \in [2, 10, 40]\%$, motivated by the range of instantaneous star formation efficiency at $\Sigma_{\rm cl} \gtrsim 3000\,M_\odot\,{\rm pc}^{-2}$ computed using hydrodynamic simulations by \citet[][their Figure 2]{gruhopfau18}. Our choice of forming $M_{\star,\rm cl}=M_{\rm cl}\epsilon_{\star}$ of stars instantaneously rather than distributing their formation through time is further motivated by the bursty star formation conditions in the primordial thick disk \citep[e.g.][]{hel20,yubulkle21}. As a point of comparison, our thick disk clouds hosting $M_{\star,\rm cl}=2{\times}10^{5}-2.4{\times}10^{8} M_{\odot}$ of stars encompass (but can exceed) the most luminous star-forming region in the Local Group, the Tarantula Nebula \citep[which hosts ${\sim}10^{6} M_{\odot}$ of stars; e.g.][]{schramtra18}.

\subsection{Stellar Dynamics}

Star clusters disperse over time at a rate that is determined by the stellar velocity dispersion $\sigma_{\star}$, which in turn affects the timescale over which stars (and protoplanetary disks) are subject to the intense radiation fields present in highly ``clustered" environments. We describe in this section how we explicitly account for the time-evolving dynamics of stellar clusters.

We assume that the initial stellar velocity dispersion is inherited from the ISM, as observed in young clusters \cite[e.g.][]{qialigol12,halixu21} and reported in simulations \citep{offhankru09}. It is well-established that the turbulent velocity dispersion of the ISM in GMCs, measured over a spatial scale $r$, scales approximately as \cite[i.e. the size-linewidth relation, e.g.][]{lar81,solrivbar87},

\begin{equation}\label{linewidth-size}
    \sigma_{\star}(r)=\sigma_{\rm cl}\bigg(\frac{r}{r_{\rm cl}}\bigg)^{1/2}.
\end{equation}
Given how $\sigma_{\rm cl}$ is higher by factors of $\sim$3--9 in star-forming environments at cosmic noon and analogous environments compared to the present-day (see Section \ref{ssec:gal-disk-gmc}), it follows from equation \ref{linewidth-size} that $\sigma_\star$ will also be higher.

While the size-linewidth relation (equation \ref{linewidth-size}) is empirical, it can also be understood from the supersonic turbulence energy spectrum. It follows that the relation holds on scales down to the sonic length, $r_{\rm s}=r_{\rm cl}(c_{\rm s}/\sigma_{\rm cl})^{2}$ (with $c_{\rm s}=0.23$ km s$^{-1}$ the GMC sound speed, assumed isothermal at 15 K; \cite{heydam15}),
below which the gas dynamics 
is governed by thermal motion rather than turbulence \citep{mur09,hop12b}. We therefore consider $r_s$ as the minimum scale of clump formation by turbulent fragmentation and set it to be the typical size of protostellar cores \citep[see also][]{offkramat10,hop12b,hop13}; i.e., this is the minimum spatial scale at which we will sample a star. 

Many young star clusters are known to exhibit ``mass segregation", in which the most massive stars are preferentially centrally located and/or have systematically lower velocity dispersions than their lower-mass counterparts. The origin of this effect may be primordial or dynamical in nature (though see e.g. \cite{trevan13} for evidence against dynamical evolution towards equipartition), but observations of young clusters indicate that mass segregation can routinely emerge from birth \cite[e.g.][]{hilhar98,kirmye11,zeisabnot21}, in agreement with some massive star formation models \cite[e.g.][]{bonbat06} and ab initio star cluster formation simulations \cite[][]{gusmaroff22}. To account for mass segregation, we additionally scale equation \ref{linewidth-size} by the factor $(m_{\star}/\big<m_{\star}\big>)^{-\beta}$ so that the full expression of $\sigma_\star(r)$ that we use is
\begin{equation}
    \sigma_\star(r) = \sigma_{\rm cl}\left(\frac{r}{r_{\rm cl}}\right)^{1/2}\left(\frac{m_\star}{\big<m_\star\big>}\right)^{-\beta}
    \label{eq:sig_star_full}
\end{equation}
where $m_\star$ is the mass of an individual star and $\big<m_{\star}\big>=0.3 M_{\odot}$ is the median stellar mass from our adopted IMF (equation \ref{equation:IMF}). We will explore $\beta\in[0,\ \ 1/2]$ where $\beta=0$ is without mass segregation and $\beta=1/2$ is chosen from the expectation of energy equipartition. We note that because the degree of primordial mass segregation is not observationally well-constrained, real $\beta$ could be higher than 1/2.

For each cluster, we draw masses of each star from our IMF (equation \ref{equation:IMF}) until the total stellar mass equals $\epsilon_\star M_{\rm cl}$. For our chosen parameters, the typical number of stars per cluster is $\sim$10$^7$. We now describe how we determine the time-varying position of each star with respect to our target star. 

The initial position of each star is set by a radial vector $\mathbf{r_{0}}$ from the subject star to a random point on the sphere of radius $r_0$ drawn from the correlation function equation \ref{equation:correlation_function}. The minimum $r_0$ is set to \texttt{max}$(r_{\rm s},\ 10^{-2} \ \rm pc)$ to ensure turbulent statistics that govern $\sigma_\star(r)$ is valid ($r > r_{\rm s}$) and to avoid contamination from binary star formation where our correlation function no longer applies (see Section \ref{ssec:clustering}). For our adopted $\Sigma_{\rm cl}$ and $\sigma_{\rm cl}$, $r_{\rm cl} = 30.6$--237.0 pc (evaluated using $\Sigma_{\rm cl} = M_{\rm cl}/\pi r_{\rm cl}^2$) and $r_{\rm s}$ is always smaller than 0.01 pc.

The velocity of each star with respect to the subject star is taken as a vector $\mathbf{v}$ on a random point on the sphere of radius $v(r_{0})$ sampled from a Maxwell-Boltzmann distribution \citep{bintre08}:

\begin{equation}\label{MB}
    p(v,r)\propto v^{2}\exp{\bigg(-\frac{v^{2}}{2\sigma^{2}_{\star}(r)}\bigg)},
\end{equation}
where $\sigma_\star(r)$ is given by equation \ref{eq:sig_star_full}. Both $\mathbf{r_0}$ and $\mathbf{v}$ are set to be constant so that at any given time $t$, stellar positions are set by 
\begin{equation}
    \mathbf{r}_{\star}(t)=\mathbf{r}_{0}+\mathbf{v}t.
    \label{eq:stellar-position}
\end{equation}
We implicitly ignore N-body accelerations in our setup, which we confirm to be valid given the relaxation time of our clusters is $\gtrsim$10 Gyr whereas our evolution only proceeds for $\lesssim$Myr, as we are primarily interested in the dissipation timescale of protoplanetary disks under external radiation. Our setup also omits the background potential from the GMC and cluster. Future work should address to what extent the background potential alters the stellar velocity distribution laid down by turbulence.

\subsection{Interstellar Attenuation}\label{ssec:attenuation}

The photoevaporative dispersal of protoplanetary disk gas is dominated by UV photons (which can be sourced from the host star \citep[e.g.][]{holjohliz94,fonmccjoh04} and/or nearby massive stars \citep[e.g.][]{johholbal98}), and X-ray emission from the host star \citep[e.g.][]{oweerccla10,oweerccla11}. X-ray emission from massive stars \citep[$L_{\rm x}{\sim}10^{-7} L_{\rm bol}$;][]{rau22} is minuscule compared to their UV output, the latter of which accounts for the bulk of their total emission, so we neglect any massive star X-ray flux. UV photons from surrounding stars, however, are subject to attenuation by the surrounding ISM en route to the target star. We outline in this section how we account for this attenuation process.

We first construct the density field using a lognormal distribution with the dispersion set by the Mach number, as expected for ISM gas under supersonic turbulence \citep[e.g.][]{vaz94,norpad99,padnor02,hop12}:

\begin{equation}\label{lognormal}
    \frac{dp}{d\delta}(\delta)=\frac{1}{\sqrt{2\pi \sigma_{\rho}^{2}(r)}}\exp{\bigg(-\frac{\delta^{2}}{2\sigma_{\rho}^{2}(r)}\bigg)},
\end{equation}

\noindent where we define the gas overdensity $\delta$ as

\begin{equation}
    \delta=\ln\bigg(\frac{\rho}{\rho_{\rm cl}}\bigg)-\bigg<\ln\bigg(\frac{\rho}{\rho_{\rm cl}}\bigg)\bigg> = \ln\bigg(\frac{\rho}{\rho_{\rm cl}}\bigg) + \frac{\sigma_{\rho}^{2}(r)}{2},
\end{equation}

\noindent from the conservation of mass and momentum and
\begin{equation}\label{turb_dispersion}
    \sigma_{\rho}(r)\approx \bigg(\ln\bigg[1+b^{2}\mathcal{M}(r)^{2}\bigg]\bigg)^{1/2},
\end{equation}
with $b$ set by the relative contribution of compressive vs.~solenoidal driving of turbulence \cite[e.g.][]{fedromkle10}, and $\mathcal{M}(r)=\sigma_{\star}(r)/c_{\rm s}$ the turbulent Mach number. We take $b=0.86$, a calculated median from the measurements of $b$ summarized in Figure 7 of \cite{shamenfed22}, which derive from a variety of star-forming environments.
The choice of $b$ should not change our result significantly given the logarithmic dependence of $\sigma_\rho$ on $b^2$.

The density field between each star and our subject star can be described by equation \ref{lognormal} but only down to $r_s$ from the subject star. Inside $r_s$ and down to the surface of the star, we consider the subject star to be embedded in a natal gas envelope (akin to the class I stage).\footnote{The inner edge of class 0/I embedding clouds is highly uncertain, depending on e.g. the presence and size of a cavity, or the coupling from envelope to disk \citep[e.g.][]{tobharcal08,chilootob12,tyclukdis21}. To be conservative, we assume the embedding envelope extends all the way to the stellar surface.} We omit the class 0 stage as class 0 protostellar envelopes are sufficiently dense to attenuate FUV photons \citep[with column number densities ${\gtrsim}10^{23}$ cm$^{-2}$, significantly larger than the FUV stopping column ${\sim}10^{21}$ cm$^{-2}$;][]{tobharloo10,huagirste24}, quenching any disk evaporation.Disks in solar neighborhood-like clouds may also be radiation-shielded by the ambient ISM early on during the first ${\sim}0.5$ Myr \citep[with a spread in timescale that depends on the GMC virial parameter;][]{qiacolhaw23,wilporcou23}. Future simulations would be helpful to elucidate how natal shielding times may differ in more turbulent, dense, and massive GMCs like those at cosmic noon. Following the turbulent accretion model of \cite{murcha15} (see also the core envelope density scaling reported in the simulations of \cite{offklemck09}), the ISM density distribution surrounding the subject star is taken to be,

\begin{equation}\label{equation:embed}
    \rho_{\rm em}(r) = \rho_{\rm em,0}\bigg(\frac{r}{r_{\rm s}}\bigg)^{-3/2}, \ r \leq r_{\rm s},
\end{equation}

\noindent 
where $r_{\rm s}$ is set to the size of the protostellar core as described earlier and $\rho_{\rm em,0}$ is drawn from equation \ref{lognormal}. From $r_{\rm s}$ to the location of each star $r_\star(t)$ at time $t$, we sample the ISM density from equation \ref{lognormal} in intervals of the sonic length.

Only the most nearby massive stars contribute to the radiation field, so we only keep track of the location of stars with $m_\star \geq 5 M_\odot$ and within 3 pc of the subject star at a given time, advancing their $\mathbf{r}_\star(t)$ (equation \ref{eq:stellar-position}) every $3{\times}10^4$ yr until $3{\times}10^6$ yr. This timestep is chosen for computational efficiency without loss of accuracy---larger timesteps cause artificial step function discontinuities in the radiation field as stellar separations expand and the radiation field drops, while smaller timesteps produce no noticeable improvement in the radiation field evolution. To account for time-varying line of sight between stars due to stellar dynamics, we redraw $\rho$ from equation \ref{lognormal} every timestep. The embedding density $\rho_{\rm em, 0}$ is re-drawn every eddy turnover time for our simulated clouds ($\sim$1.7-4.8 Myr). Note that we do not account for any embedded structure around massive stars because they are expected to establish HII regions over a timescale ${\sim} (\alpha_{\rm b}n_{\rm H})^{-1}{\sim} 10^{2}$ yr before undergoing rapid hydrodynamic expansion down the density gradient into the surrounding ISM \cite[e.g.][]{fratenbod90,bishawwil15,zamvazgon19}.

Since EUV ($h\nu =13.6-124$ eV) and FUV ($h\nu=6-13.6$ eV) photons can both influence protoplanetary disk lifetimes, we track the time evolution of both radiation fields. We first define $Q_{\star}$ as the number of photons s$^{-1}$ emanating from the surface of a given massive star. The photon flux at some radius $r^{\prime}$ from the massive star is,

\begin{equation}\label{equation:flux}
    \Phi(r^{\prime})=\frac{Q_{\star}}{4\pi r^{{\prime}^{2}}}e^{-\tau(r^{\prime})}
\end{equation}

\noindent where $\tau(r^{\prime})$ is the optical depth along the path. 

For EUV photons, we account for attenuation by both the dust and the ionization of hydrogen following \citep{dra11},

\begin{equation}\label{equation:tau_euv}
    \tau_{\rm EUV}(r^{\prime})=\int_{0}^{r^{\prime}} n_{\rm H}(R)\bigg[(1-x_{i}(R))\sigma_{\rm EUV}+\kappa_{\rm d, \rm EUV}\eta_{\rm d} m_{\rm H}\bigg] dR,
\end{equation}

\noindent where $n_{\rm H}$ is the number density of all hydrogen species, $x_i \equiv n_{\rm H^+}/n_{\rm H}$ is the ionization fraction of hydrogen, $n_{\rm H^+}$ is the number density of ionized hydrogen, $\sigma_{\rm EUV} = 2\times10^{-18} \ \textrm{cm}^{2}$ is the hydrogen ionization cross section to EUV photons of 20 eV \citep{spi78}, $\kappa_{\rm d, \rm EUV}=2{\times}10^{4} \ \textrm{cm}^{2} \ \textrm{g}^{-1}$ is the opacity to EUV photons per gram of dust \citep[][]{kuiyormig20}, $\eta_{\rm d}=10^{-2}$ is the dust-to-gas ratio, and $m_{\rm H}$ is the mass of a hydrogen atom.\footnote{Using a photoionization cross section that varies with EUV wavelength \citep[e.g.][]{verferkor96}, and therefore stellar effective temperature, has a minor effect on our results. Including this physics changes the EUV photon flux across the disk surface by at most ${\lesssim}30\%$, and by ${\lesssim}5\%$ at the disk outer edge throughout the entire evolution. Our EUV evaporation rates therefore deviate by ${\lesssim}14\%$, and would not significantly alter our disk lifetime estimates.} We compute $x_{i}$ at each point in space following \citet{dra11} according to ionization-recombination balance:

\begin{equation}\label{eq:x_i}
        x_{\rm i}(r') = \frac{1}{2}\frac{\Phi(r')\sigma_{\rm EUV}}{\alpha_{\rm b}n_{\rm H}(r')}\left[-1+\sqrt{1+4\left(\frac{\Phi(r')\sigma_{\rm EUV}}{\alpha_{\rm b}n_{\rm H}(r')}\right)^{-1}}\right]
\end{equation}

\noindent with $\alpha_{\rm b}=2.54\times10^{-13} \ \textrm{cm}^{3} \ \textrm{s}^{-1}$ the case b recombination coefficient for 10$^{4}$ K ionized hydrogen \citep[][]{dra11}, 
to account for re-ionization of the ground state. For simplicity, we assume the ISM is pure hydrogen. We calculate the radiation field impinging on our subject star by evaluating equations \ref{equation:flux}--\ref{eq:x_i} at $r^{\prime}=r_{\star}$.

For FUV radiation, we only account for dust attenuation,

\begin{equation}\label{equation:tau_fuv}
    \tau_{\rm FUV}(r^{\prime})=\int_{0}^{r^{\prime}} n_{\rm H}(R)\kappa_{\rm d, \rm FUV}\eta_{\rm d} m_{\rm H} dR,
\end{equation}

\noindent with $\kappa_{\rm d, \rm FUV}=4{\times}10^{4} \ \textrm{cm}^{2} \ \textrm{g}^{-1}$ the opacity to FUV per gram of dust \citep[][]{kuiyormig20}.

We calculate the stellar EUV and FUV fluxes as functions of effective temperature $T_{\rm eff}$ and surface gravity $\log g$ from the stellar atmosphere models of \cite{caskur03},\footnote{Available here: \url{https://www.stsci.edu/hst/instrumentation/reference-data-for-calibration-and-tools/astronomical-catalogs/castelli-and-kurucz-atlas}} with a stellar metallicity [Fe/H]=$-0.5$, appropriate for thick disk stars. Each pair of $T_{\rm eff}$ and $\log g$ is mapped to a stellar mass and age using the MIST stellar evolution tracks with [Fe/H]=$-0.5$ and rotation speeds $40\%$ of breakup \cite[][]{paxbildot11,paxcanarr13,paxmarsch15,chodotcon16,dot16}.\footnote{Available here: \url{https://waps.cfa.harvard.edu/MIST/}} 
To obtain total EUV and FUV fluxes, we integrate the photon fluxes over 100-911 $\AA$ ($h\nu=13.6-124$ eV), and 911-2066 $\AA$ ($6-13.6$ eV), respectively. 
We then convert these fluxes $F_{\rm EUV/FUV}(\log g(m_{\star}, t), T_{\rm eff}(m_{\star}, t))$ to photon luminosities as $Q_{\star,\rm EUV/FUV} \equiv F_{\rm EUV/FUV}{\times}4\pi R^{2}_\star(m_{\star},t) / h\nu_{\rm EUV/FUV}$ where we use the MIST grid for the stellar radius $R_\star$, and we take the photon energies $h\nu_{\rm EUV} = 20$eV and $h\nu_{\rm FUV}$ = 9.75 eV where $h$ is the Planck constant and $\nu_{\rm EUV/FUV}$ are representative EUV/FUV photon frequencies. We find that massive stars' $\log g$ and $T_{\rm eff}$ frequently fall outside the range of the \cite{caskur03} atmosphere models during their pre-main sequence contraction phase. We therefore begin our UV luminosity track calculations at $10^{4}$ yr, at the beginning of the massive stars' main sequence evolution. We also find that the most massive stars ($\gtrsim 70 M_{\odot}$) have slightly larger $T_{\rm eff}$ than the maximum covered by \cite{caskur03} (${\sim}5.3{\times}10^{4}$ vs. $5{\times}10^{4}$ K), so we evaluate their UV luminosities assuming $T_{\rm eff}=5{\times}10^{4}$ K.

For each set of cloud parameters, we run $10^{2}$ Monte Carlo iterations of our calculation, including all steps from sections \ref{ssec:clustering} to \ref{ssec:attenuation}. We then take the 50th percentile output for the EUV, FUV, and bolometric fluxes. This is necessary because randomly sampling stellar positions from the correlation function produces fluctuations in the radiation field. We find that the median and standard deviation of UV flux for a given cloud converge at $10^{2}$ iterations.

\subsection{Protoplanetary Disk Evolution}

Given an impinging radiation field, we calculate the 
%EJL
%secular 
one-dimensional
evolution of a protoplanetary disk under viscous diffusion, heating, and evaporative mass loss. Our disks undergo bolometric heating from the ambient cluster and the central star, internal dissipation, and ionization heating from cosmic rays. We track two contributions to disk evaporation: an internally catalyzed X-ray wind, and an FUV or EUV-driven wind catalyzed by nearby massive stars. Each contribution is described in detail in the following sections.

Following \citet{lynpri74} and \citet{pri81}, additionally accounting for the mass loss by wind,

\begin{align}\label{equation:viscous}
\begin{split}
    \frac{\partial{\Sigma(R,t)}}{\partial t}&=\frac{3}{R}\frac{\partial}{\partial R}\bigg[R^{1/2}\frac{\partial}{\partial R}(R^{1/2}\nu(R,t)\Sigma(R,t))\bigg] \\
    &- \dot{\Sigma}(R,t)_{\rm wind},
\end{split}
\end{align}

\noindent where $\Sigma$ is the disk gas surface density, $R$ is the disk radial coordinate, $\nu \equiv \alpha c_s H$ is the viscosity, $\alpha$ is the Shakura-Sunyaev viscosity parameter \citep[][]{shasun73} which we fix to $10^{-3}$, $c_{\rm s}(R)=\sqrt{k_{\rm B}T(R)/\mu}$ is the disk gas sound speed, $k_{\rm B}$ is Boltzmann's constant, $T(R)$ is the disk gas midplane temperature, $\mu=2.33$m$_{\rm H}$ is the gas mean molecular weight, $H=c_{\rm s}/\Omega$ is the disk gas scale height, and $\Omega$ is the Keplerian orbital frequency. The mass loss rate by wind $\dot{\Sigma}_{\rm wind}=\dot{\Sigma}_{\rm int}+\dot{\Sigma}_{\rm ext}$ contains energy sources from the central star's X-ray and the UV rays from the external radiation field, in that order.

%EJL
%To solve equation \ref{equation:viscous}, we change variables. Setting $y\equiv 2 R^{1/2}$ and $z\equiv y\Sigma$ yields \citep[][]{batpri81},
We rewrite equation \ref{equation:viscous} with $y\equiv 2 R^{1/2}$ and $z\equiv y\Sigma$ \citep[][]{batpri81},
\begin{equation}\label{equation:viscous_cov}
    \frac{\partial z}{\partial t}=\frac{12}{y^{2}}\frac{\partial^{2} (z\nu)}{\partial y^{2}} - \dot{z}_{\rm wind}.
\end{equation}

\noindent Equation \ref{equation:viscous_cov} is solved implicitly with a backwards-Euler scheme. We use a grid of 2000 cells equally spaced in $R^{1/2}$ from 0.1 to 100 au. We employ a torque-free inner boundary condition ($2\pi \nu\Sigma R^{3}d\Omega/dr=0$), and an outflow outer boundary condition ($\partial/\partial R (2\pi \nu\Sigma R^{3}d\Omega/dr)=0$).
%EJL
%The outer boundary is however unused for our disks that evaporate.
For stars in thick disk that we find to undergo rapid evaporation, the outer boundary condition is controlled by the evaporative mass loss and so we find negligible difference whether or not we set the aforementioned explicit outer boundary condition.
%We have verified that using an outflow inner boundary condition does not change the evolution. 
%
We set the timestep $\delta t=$10 yr and have verified that our results are converged in time and space. The initial condition for our disk is taken from the zero-time self-similar solution of \cite{lynpri74},

\begin{equation}\label{equation:IC}
\Sigma(R,0)\propto \bigg(\frac{R}{R_{1}}\bigg)^{-\gamma}\exp{\big[-\bigg(\frac{R}{R_{1}}\bigg)^{2-\gamma}\big]},
\end{equation}

\noindent with $\nu\propto R^{\gamma}$ so that $\gamma=3/2-\beta$ if $T\propto R^{-\beta}$. Our 
%EJL
%disk evolution 
%
calculations account for 
%EJL
%time 
temporally
and spatially varying midplane temperature (see Section \ref{sec:disk_temp}). We find a posteriori that our protoplanetary disks at cosmic noon are so strongly irradiated by the ambient cluster that they are isothermal beyond $\gtrsim$10 au, 
so we set $\gamma=3/2$ as an initial guess (ultimately, inside $\sim$10 au, the disk evolves away from isothermal due mainly to viscous dissipation). We initialize our disk simulations in the solar neighborhood with $\gamma=1$, as we find that heating from the central star dominates the temperature profile beyond $\sim$5--10 au. We have verified that our disk evolution simulations are insensitive to whether we set $\gamma$=1 or 3/2 in our initial conditions. 
%EJL
%Using equation \ref{equation:IC} is therefore not expected to strongly affect our results, despite the fact that it is derived for disks with temporally constant $\nu(R)$ \citep{lynpri74}. 
We therefore expect our results to be robust against our choice of initial $\Sigma$ (equation \ref{equation:IC} which is derived assuming temporally constant $\nu(R)$ \citep{lynpri74}).
Equation \ref{equation:IC} is normalized with an initial disk mass M$_{\rm d}(0)=0.1 M_{\star}$, with $M_{\star}=1 M_{\odot}$ being the host stellar mass and $R_{1}$=40 au is the characteristic radius. We have varied $R_1\in [10,\ 60]$ au and find very little variation in our results. 

Throughout this paper, we define the disk outer radius $R_{\rm disk}$ at the outermost cell with $\Sigma$ above $10^{-4}$ g cm$^{-2}$, roughly corresponding to the outermost cell containing non-zero mass.

\subsubsection{Disk Temperature}\label{sec:disk_temp}

We compute the midplane temperature of the disk following \cite{naknak94} (see also \cite{sirgoo03}), with a $T(\tau)$ solution that interpolates between the optically thick and thin regimes:

\begin{align}\label{equation:disk_temp}
\begin{split}
    \sigma_{\rm SB} T^{4} &= F_{\rm vis}\bigg[\frac{1}{2}+\frac{3}{16}\tau_{\rm R}+\frac{1}{4\tau_{\rm p}}\bigg]+F_{\rm cl} \\
                          &+ F_{\rm CR}\bigg[\frac{1}{2}+\frac{1}{4\tau_{\rm p}}\bigg]+F_{\star},
\end{split}
\end{align}

\noindent where $\sigma_{\rm SB}$ is the Stefan-Boltzmann constant, $F_{\rm vis}$ is the viscous dissipation rate, $\tau_{i}=\kappa_{i}\Sigma$ is the Rosseland ($i=$R) or Planck ($i=$p) mean optical depth with $\kappa_{i}$ the Rosseland/Planck mean opacity, $F_{\rm cl}$ the bolometric flux from the ambient cluster, $F_{\rm CR}$ the cosmic ray heating rate, and $F_{\star}$ the flux from the host star. We use the opacities of \cite{semhenhel03}  (for iron-poor, porous composite, spherical dust grains) to approximate $\kappa_{\rm R} = 5$ cm$^{2}$ g$^{-1}$ (per gram of dust and gas mixture) when $T\geq 150$ K, $\kappa_{\rm R}\propto T^{2}$ when $T<150$ K, and $\kappa_{\rm p}\approx 1.1\kappa_{\rm R}$ (see also \cite{naknak94,tho13}).

The total energy generation rate per unit area from viscous dissipation is $F_{\rm vis}=\nu\Sigma(R \partial\Omega/\partial R)^{2}$ \citep[][]{frakinrai02}, giving,

\begin{equation}\label{equation:vis_dis}
    F_{\rm vis}=\frac{9}{4}\nu\Sigma\Omega^{2}
\end{equation}
which radiate through both sides of the disk so that the contribution to the temperature purely due to $F_{\rm vis}$ is $(1/2 F_{\rm vis} \sigma_{\rm SB}^{-1})^{1/4}$ in the optically thin limit (c.f.~equation \ref{equation:disk_temp}).

We follow equation 4 of \citet{dalcancal98} to write the cosmic ray heating rate, integrated over both disk faces:

\begin{equation}\label{equation:cr}
    F_{\rm CR}=\frac{2}{2 m_{\rm H}}\Delta Q_{\rm CR}\Gamma_{\rm CR}\zeta_{\rm CR,0}\lambda(1-e^{-\Sigma/2\lambda}),
\end{equation}

\noindent where $\Delta Q_{\rm CR}=20$ eV \citep{gollan78} is the energy liberated in heating during each ionization, $\zeta_{\rm CR,0}=5{\times}10^{-17}$ s$^{-1}$ is the present-day Galactic cosmic ray ionization rate per H$_2$ molecule \cite[][]{web98,vanvan00}, $\Gamma_{\rm CR}$ is a flux enhancement factor over $\zeta_{\rm CR,0}$ which we take as 1 for solar neighborhood and $10^4$ for the thick disk conditions, adopting the values for starburst environments \citep[e.g.,][]{pap10}, and $\lambda=96$ g cm$^{-2}$ is the cosmic ray attenuation surface density \citep{dalcancal98}. We find a posteriori that our final conclusion is robust against the exact value of $\Gamma_{\rm CR}$ we use. Given the relatively small attenuation column depth, we have assumed the cosmic ray heating is deposited at the surface of the disk in writing equation \ref{equation:disk_temp}. 

The stellar heating flux for a flared disk is \cite[e.g.][]{rudpol91,huegui05},

\begin{equation}\label{equation:fstar}
    F_{\star}=\sigma_{\rm SB}T^{4}_{\star}\bigg[\frac{2}{3\pi}\bigg(\frac{R_{\star}}{R}\bigg)^{3}+\frac{1}{2}\bigg(\frac{R_{\star}}{R}\bigg)^{2}\frac{H}{R}\bigg(\frac{d\ln{H}}{d\ln{R}}-1\bigg)\bigg],
\end{equation}

\noindent where the first term reflects the flux impinging on a flat disk, the second corrects for a non-zero flaring angle of the disk, and $T_{\star}$ and $R_{\star}$ are the central star's effective temperature and radius, respectively, evolved using the MIST evolution tracks for a 1 $M_{\odot}$ star. In order to avoid a known convergence issues \citep{huegui05}, we fix $d\ln H / d\ln R$ instead of evolving them numerically. Beyond $\sim$10 AU where the flaring term matters, we find $T$ to be radially constant in the thick disk environment (so that $d\ln H/d\ln R = 3/2$) whereas $T \propto R^{-1/2}$ in the solar neighborhood (so that $d\ln H / d\ln R = 5/4$) and so we set $d\ln H / d\ln R$ to these constants accordingly. At smaller orbital distances, $d\ln H /d\ln R$ is no longer constant but other heating sources (viscous dissipation in both thin and thick disk, with additional contributions from cosmic ray heating in the thick disk) dominate.

Equation \ref{equation:disk_temp} is solved iteratively at each timestep. We first set $T(R)$ according to equation \ref{equation:disk_temp}, then update the viscous flux (which depends on $T$ through $\nu$) and stellar heating flux (which depends on $T$ through $H$), before updating $T(R)$ once more. We repeat this procedure of updating the fluxes followed by the temperature until $T(R)$ does not change between iterations by more than 1 K; this typically requires ${\sim}$2 rounds of iteration per timestep. Above temperatures $\sim$1500 K, dust vaporizes and the disk opacity drops, maintaining an approximately isothermal temperature \citep{dalcancal98}. We therefore enforce a maximum temperature of 1500 K instead of solving for the drop in opacity due to silicate vaporization.

\subsubsection{Disk Accretion \texorpdfstring{$\&$}{and} Evaporation}

At each radius $R$ and time $t$, we compute the viscous mass radial flux 

\begin{equation}\label{equation:mdot_disk}
    \dot{m}_{\rm vis}(R,t)=2\pi R v_{R}(R,t) \Sigma(R,t),
\end{equation}

\noindent with the gas radial velocity given by \citep{frakinrai02}

\begin{equation}
v_{R}(R,t)=-\frac{3}{\Sigma R^{1/2}}\frac{\partial}{\partial R}(R^{1/2}\nu(R,t)\Sigma(R,t)).
\end{equation}
Mass flow through the inner boundary 0.1 au is taken as accretion onto the central star $\dot{m}_{\star}=\dot{m}_{\rm vis}(0.1 \ \mathrm{au}, t)$.

Next, we employ the X-ray wind mass loss profile $\dot{\Sigma}(R)$ and mass loss rate provided in Table 2 of \cite{pigercesp21}, and the mass loss rate/X-ray luminosity scaling from \cite{picercowe19}, appropriate for a 1 $M_\odot$ central star. Taking the 50th percentile stellar X-ray luminosity from Figure 1 of \cite{oweerccla11}, which contains X-ray luminosity functions for 0.5-1$M_{\odot}$ stars from Taurus \citep{gudbriarz07} and Orion \citep{prekimfav05}, we fix $L_{\rm X}=10^{30}$ erg s$^{-1}$. These choices yield a total mass loss rate of $\dot{m}_{\rm X}=2.4{\times}10^{-8} \ M_{\odot } \ \rm{yr}^{-1}$.

Evaporation by external radiation can be EUV or FUV-dominated. As a brief summary of the underlying physics, EUV photons control the outflow when the EUV ionization front lies sufficiently close to the disk that the FUV-heated gas launched from the disk surface cannot pass through a sonic point. Following \cite{hawcolqia23}, we approximate the transition between EUV/FUV-driven mass loss by taking $\dot{m}_{\rm wind}=\texttt{max}(\dot{m}_{\rm EUV},\ \dot{m}_{\rm FUV})$, where $\dot{m}_{\rm EUV}$ is the EUV-driven mass loss rate and $\dot{m}_{\rm FUV}$ is the FUV-driven mass loss rate. We stop the disk evolution calculations when the disk mass is $10^{-5} M_{\odot}$. In the rest of this section, we will describe how we solve for the EUV and FUV-driven mass loss rates, and how we implement external evaporation.

The EUV mass loss rate is evaluated at the disk outer edge, $R_{\rm disk}$.\footnote{Using three orders of magnitude larger or smaller thresholds in $\Sigma$ to define $R_{\rm disk}$ than our fiducial $10^{-4}$ g cm$^{-2}$ changes disk lifetimes by ${<}10^{5}$ yr.} This choice is supported by numerical simulations that find most of the material is removed from the disk outer edge \citep[][]{ricyor98,ricyor00}, and semi-analytic modelling which compares well to observed disk sizes and mass loss rates \citep{johholbal98,stohol99}. Following \cite{johholbal98},

\begin{align}\label{equation:mdot_euv}
\begin{split}
    \dot{m}_{\rm EUV}(R_{\rm disk}) &=\\
    & \hspace*{-1.6cm} 4\pi\epsilon_{\rm EUV}\mathcal{F}\sqrt{\frac{3}{\alpha_{\rm b}}} \mu_{\rm EUV} v_{\rm EUV} \Phi^{1/2}_{\rm EUV}(R_{\rm disk})(1+x_{\rm PDR})^{3/2}R^{3/2}_{\rm disk} \\
    & \hspace*{-1.6cm} \approx  2{\times}10^{-7} \bigg(\frac{R_{\rm disk}}{100 \ \mathrm{au}}\bigg)^{3/2}\bigg(\frac{\Phi_{\rm EUV}(R_{\rm disk})}{10^{13}\ \mathrm{cm}^{-2} \ \mathrm{s}^{-1}}\bigg)^{1/2}\\
    & \ M_{\odot} \ \mathrm{yr}^{-1},
\end{split}
\end{align}

\noindent where $\epsilon_{\rm EUV} = 1.86$ is a numerical efficiency factor \citep{johholbal98}, $\mathcal{F}$ a geometric correction for the fraction of the solid angle subtended by the wind, $\mu_{\rm EUV} = 0.5 m_{\rm H}$ is the mean molecular weight in the EUV region, $v_{\rm EUV}\approx c_{\rm s, EUV} = 10 \ \mathrm{km \ s^{-1}}$ the terminal outflow velocity at the EUV ionization front, appropriate for an ionized gas at $\sim$10$^4$ K, $\Phi_{\rm EUV}$ the impinging EUV flux, and $x_{\rm PDR} \equiv r_{\rm PDR}/R_{\rm disk}$ the dimensionless radial extent of the photodissociation region (PDR). Our adopted $\epsilon_{\rm EUV}$ is appropriate for $x_{\rm PDR} \ll 1$.

We solve for $x_{\rm PDR}$ following ionization-recombination balance throughout the EUV portion of the wind by taking the positive root of equation 22 of \cite{johholbal98}:

\begin{equation}\label{equation:xpdr}
\frac{x_{\rm PDR}^{2}}{1+x_{\rm PDR}}=\frac{\alpha_{\rm b}}{3}\frac{N_{\rm FUV}^{2}}{R_{\rm disk}}\bigg(\frac{\mu_{\rm FUV}}{\mu_{\rm EUV}}\frac{c^{2}_{\rm s,\rm FUV}}{2 c_{\rm s, \rm EUV} v_{\rm EUV}}\bigg)^{2}\Phi^{-1}_{\rm EUV}(R_{\rm disk}),
\end{equation}

\noindent where $N_{\rm FUV}\approx 10^{21}$ cm$^{-2}$ is the PDR column density chosen so that the FUV optical depth is $\sim$1 \citep[e.g.][]{johholbal98,adahollau04}, $\mu_{\rm FUV}=m_{\rm H}$ is the mean molecular weight of FUV outflow, appropriate for a neutral gas, and $c_{\rm s, FUV} = 3\,{\rm km \ s^{-1}}$ is the sound speed of FUV outflow. We enforce a maximum $x_{\rm PDR}=1.5$, appropriate for an EUV-driven outflow \citep{johholbal98}.

Assuming the EUV outflow is spherically symmetric, $\mathcal{F} = 1$, valid for flows that escape the star's potential; i.e. where $R > GM_\star / c^2_{\rm s, EUV}$ which is $\sim$9 au for our chosen parameters (numerical simulations of EUV-driven winds would be helpful to confirm $\mathcal{F}{\sim}1$, as uncertainties in $\mathcal{F}$ directly affect the mass loss rate). We therefore employ equations \ref{equation:mdot_euv} and \ref{equation:xpdr} with $\mathcal{F}=1$ for $R_{\rm disk}>GM_\star / c^2_{\rm s, EUV}$. Inside this EUV Bondi radius, we adopt the empirical fitting to the mass loss rates calculated by \citet[][their Appendix B]{owelin23} who find outflows can still be launched from inside the EUV Bondi radius. We find in the vast majority of parameter space $x_{\rm PDR}\ll 1$ when applying equation \ref{equation:mdot_euv}, validating our choice of $\epsilon_{\rm EUV}$. Cases when $x_{\rm PDR}{\sim}1$ only occur when the mass loss transitions to being FUV-driven (e.g., $\Phi_{\rm EUV}\rightarrow 0$).\footnote{There is some ambiguity in when to decide between FUV and EUV-driven evaporation. \cite{johholbal98} argue that EUV dominates when the ionization front is close enough to the disk that the PDR cannot pass through a sonic point, i.e. $x_{\rm PDR}\lesssim 1.5$. Our approach ($\dot{m}_{\rm wind}=\texttt{max}(\dot{m}_{\rm EUV},\ \dot{m}_{\rm FUV})$, as suggested by \cite{hawcolqia23}), generally agrees with \cite{johholbal98}. There are instances however when this approach produces FUV-driven evaporation even when the \cite{johholbal98} formalism predicts it should be EUV-driven ($\dot{m}_{\rm FUV}>\dot{m}_{\rm EUV}$, when $x_{\rm PDR}\ll 1$; e.g. during the first $\sim$10$^{3}$ yr of the simulation in the upper right panel of Figure \ref{figure:mdot_paramspace_eff}). We experimented with enforcing $\dot{m}_{\rm wind}=\dot{m}_{\rm EUV}$ when $x_{\rm PDR}<1.5$, and $\dot{m}_{\rm wind}=\dot{m}_{\rm FUV}$ when $x_{\rm PDR}\geq1.5$ and found no significant variation in our results.}

\begin{deluxetable*}{CCCCCCc}\label{param_table}
\tablecaption{Adopted model parameters.
\label{tab:parameters}}
\tablecolumns{6}
\tablewidth{0pt}
\tablehead{
\colhead{Parameter} &
\colhead{Definition} &
\colhead{Values (Solar Neighborhood)} & 
\colhead{Values (Thick Disk)} & 
\colhead{Units} & 
\colhead{References} &
}
\startdata
$\sigma_{\rm gal}$ & \rm \ galactic \ disk \ ISM \ velocity \ dispersion \ (log_{10}) & \nodata & [1.4, \ 1.66, \ 1.83] & \mathrm{km} \ \mathrm{s}$^{-1}$ & (1) \\
$\Sigma_{\rm gal}$ & \rm galactic \ disk \ ISM \ surface \ density & \nodata & 1700 & M_{\odot} \ \mathrm{pc}$^{-2}$ & (2) \\
$M_{\rm cl}$ & \rm GMC \ mass & 10^{5} & [10^{7}, \ 10^{8}, \ 6{\times}10^{8}] & M_{\odot} & (3), \ (4) \\
$\Sigma_{\rm cl}$ & \rm GMC \ surface \ density & 60 & 3400 & M_{\odot} \  \mathrm{pc}^{-2} & (3), \ (4) \\
$\epsilon_{\star}$ & \rm star \ formation \ efficiency & 2$\%$ & [2, \ 10,\ 40]$\%$ & \nodata & (5) \\
$\beta$ & \rm degree\ of \ mass \ segregation & 0 & [0, \ 1/2] & \nodata & \nodata \\
$\Gamma_{\rm CR}$ & \rm cosmic \ ray \ flux\ enhancement\ factor & 1 & 10^{4} & \nodata & (6) \\
$L_{\rm X}$ & \rm host \ stellar \ X-ray \ luminosity & 10^{30} & 10^{30} & \mathrm{erg} \ \mathrm{s}^{-1} & (7) \\
$\dot{m}_{\rm X}$ & \rm internal \ X-ray \ wind \ mass \ loss \ rate & 2.4{\times}10^{-8} & 2.4{\times}10^{-8} & M_{\odot} \  \mathrm{yr}^{-1} & (8), \ (9) \\
$M_{\star}$ & \rm host \ stellar \ mass & 1 & 1 & M_{\odot} & \nodata \\
M_${\rm d}(0)$ & \rm protoplanetary \ disk \ initial \ mass & 0.1 & 0.1 & $M_{\star}$ & \nodata \\
$R_{1}$ & \rm protoplanetary \ disk \ initial \ characteristic \ radius & 40 & 40 & \mathrm{au} & \nodata \\
$\gamma$ & $\rm disk \ initial \ viscosity \ power \ law \ index \ (\nu \propto R^{\gamma})$ & 1 & 1.5 & \nodata & \nodata \\
\enddata
\tablerefs{(1) \cite{pilnelspr19,girfisbol21}, (2) \cite{tacnergen13}, (3) \cite{mivmurlee17}, (4) \cite{swipapcox11,swidyenig15,desriccom19,desriccom23}, (5) \cite{gruhopfau18}, (6) \cite{pap10}, (7) \cite{oweerccla11}, (8) \cite{picercowe19}, (9) \cite{pigercesp21} }
\end{deluxetable*}

Next, we employ $\dot{m}_{\rm FUV}$ calculated by \cite{hawclarah18}, as compiled in their \texttt{FRIED} grid. The \texttt{FRIED} grid extends across FUV photon fluxes $F_{\rm FUV} \in [10, 10^{2}, 10^{3}, 5{\times}10^{3}, 10^{4}]$ G0 (with G0=$1.6{\times}10^{-3}$ erg cm$^{-2}$ s$^{-1}$ the Habing unit of UV radiation in the FUV band), disk radii $R = 1$--$400$ au, and disk masses that range from 0.0032\% to 20\% of the stellar mass. We use their 1$M_{\odot}$ mass loss rates. Although our simulated disk extends from 0.1 to 100 au, we find that our disks are destroyed (mass $<10^{-5} M_{\odot}$) before their outer radii are truncated inside 1 au, justifying our use of \texttt{FRIED} grid; in fact, the truncated inner disk is expected to rapidly drain onto the star under viscous dissipation \citep{oweerccla11}. 
Following \cite{selboocla20}, we calculate $\dot{m}_{\rm FUV}$ at the location in the disk where the flow transitions from optically thick to thin in the FUV. The FUV mass-loss rate rises outward in the optically thick region as the gas becomes more weakly bound, whereas it decreases outward in the outer optically thin region, scaling linearly with the gas surface density \citep{facclabis16}. It follows that the actual 
%EJL
%moss 
mass
loss rate is where $\dot{m}_{\rm FUV}$ attains a maximum. We find this maximal mass loss rate by computing $\dot{m}_{\rm FUV}$ at each radius in the disk as if it were the disk outer edge, interpolating \texttt{FRIED} over radius, disk mass unit $M_{400}=2\pi \Sigma_{\rm out} R^{2}_{\rm d} \big(R/400 \ \rm au \big)^{-1}$, and $F_{\rm FUV}$. \footnote{We have verified that our approach of computing $F_{\rm FUV}$ locally produces very similar results to using a larger $F_{\rm FUV}$ evaluated further out in the wind at ``infinity" ($10^{3}$ au in \texttt{FRIED}; \cite{hawclarah18}). Our locally calculated $F_{\rm FUV}{>}10^{4}$ G0 across the protoplanetary disk for the vast majority of evolution at cosmic noon, and because the FUV mass loss rate is expected to plateau above ${\gtrsim}10^{4}$ G0 \citep{winhaw22}, we find no difference in $\dot{m}_{\rm FUV}$ when using local fluxes or that further out in the wind. We have verified that using $F_{\rm FUV}$ as computed at 1000 au in the solar neighborhood does not change our disk lifetime since the disk dissipation remains controlled by the X-ray flux from the central star.}

We implement external evaporation by removing mass outside-in, in keeping with numerical simulations \citep[e.g.][]{ricyor00,adahollau04,hawcla19} and observations \citep[e.g.][]{henart98,eisarcbal18} of EUV and FUV-evaporating disks. Given the mass loss rate $\dot{m}_{\rm wind}$, we compute the total mass lost at each time step $\delta m_{\rm wind}=\dot{m}_{\rm wind}\delta t$ and remove mass from each cell starting from the disk outer edge until the total mass lost equals that from external evaporation. We remove the total mass in each cell until we reach the innermost cell where only a fraction of its total mass is removed. We have experimented with other schemes in which the mass loss is more distributed, e.g. using $\dot{\Sigma}(R)\propto R^{-1/2}$, which follows dimensionally from assuming ionization-recombination equilibrium at each radius. The evolution calculations are not sensitive to these details, generally differing by ${\sim}10^{4}$ yr, and at most by ${\sim}10^{5}$ yr, in total disk lifetime.

\begin{figure*}
\centering
\includegraphics[width=1\textwidth]{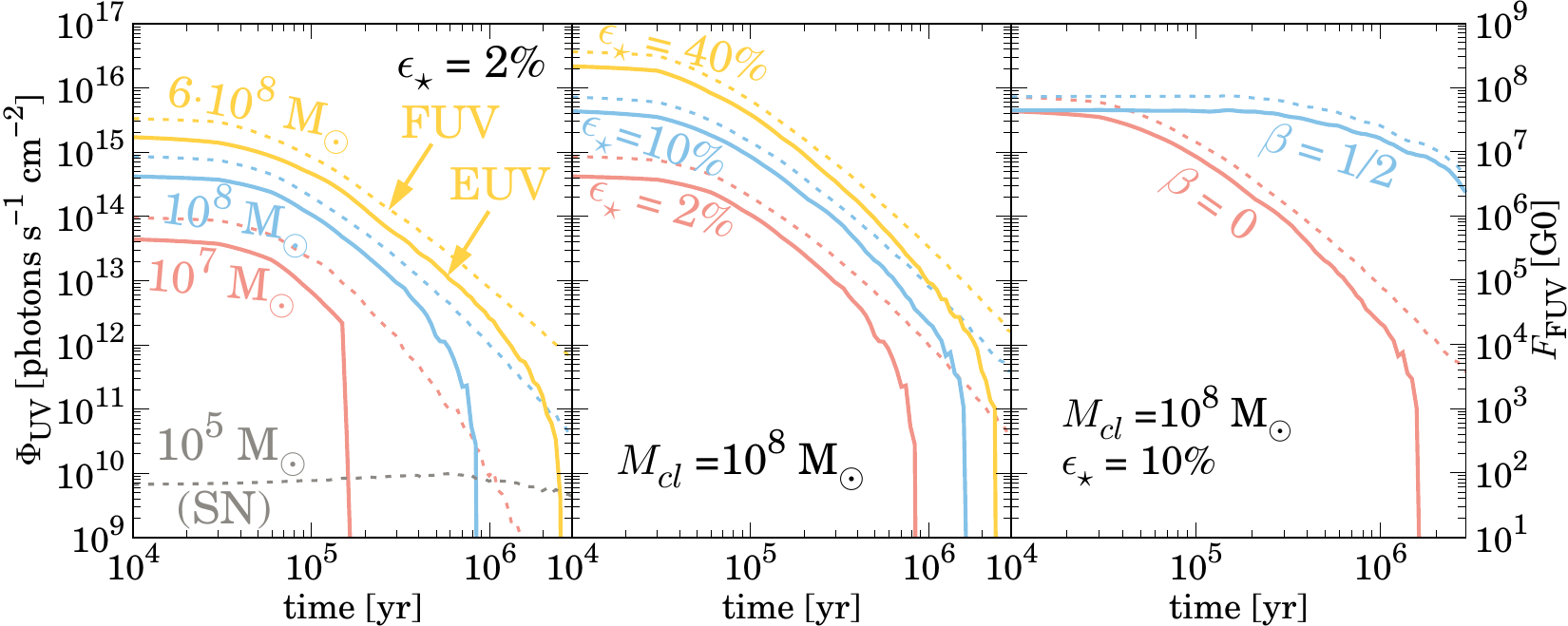}
\caption{Evolution tracks of UV photon flux $\Phi_{\rm UV}$ impinging at 100 au from a random star in various ``clustered" stellar environments. EUV fluxes are depicted in solid lines, and FUV are shown in dashed. The rightmost vertical axis enables conversion of FUV flux to Habing units (G0) for reference. Left: the time evolution of UV flux in giant molecular clouds of differing total masses at fixed star formation efficiency $\epsilon_{\star}=2\%$. Solar neighborhood condition is drawn in gray, labeled as ``SN", as a reference. Middle: UV flux for a cloud of $10^8\,M_\odot$ for varying $\epsilon_\star$. Right: variation in UV flux for fixed cloud mass and $\epsilon_\star$ with respect to the initial stellar velocity distribution (mass-segregated in blue; no mass segregation in red). In general, we find larger UV flux for larger cloud mass, higher $\epsilon_\star$, and in clouds whose stars follow an initial velocity dispersion that is mass dependent. In all cases, the clouds born in the primordial thick disk are under $\Phi_{\rm UV}$ that exceed the solar neighborhood condition by multiple orders of magnitude.
\label{figure:euvt}}
\end{figure*} 

\section{Results}\label{sec:results}

\subsection{``Clustered" Irradiation Environments}

In Figure \ref{figure:euvt} we plot evolution tracks for the UV flux impinging at 100 au from a subject star in our various environments. The left panel depicts the dependence on GMC mass with the conditions of the solar neighborhood drawn as a reference. We observe that, at early times, the solar neighborhood cloud produces $\sim$4--6 orders of magnitude lower FUV flux (and no EUV flux) than clouds in the thick disk. The FUV flux in the solar neighborhood cloud is ${\sim}70$ G0, which compares well to empirical estimates from the Orion A and B clouds \citep{vanhac23}, which have similar masses to our solar neighborhood cloud \citep[${\sim}10^{5} M_{\odot}$;][]{bal08}. This flux falls on the upper tail of the distributions of maximally extincted FUV flux (appropriate for comparison to our model, which also includes significant extinction) from \cite{fatada08} and that of \cite{winkruche20} for systems of similar bulk density ${\sim}10^{-2} M_{\odot}$ pc$^{-2}$. Our value exceeds the peaks of these distributions (at ${\sim}0.35$ G0 in \cite{fatada08}, and ${\sim}1$ G0 in \cite{winkruche20}) due to the effect of stellar clustering which produces a smaller minimum distance between the subject star and the nearest massive star. The lower flux in the solar neighborhood than in the thick disk is a result of the more than two orders of magnitude lower stellar mass in the solar neighborhood cloud, coupled with the larger sonic length in the solar neighborhood compared to the thick disk ($r_{\rm s}{\sim}0.3$ vs. $ 5\times 10^{-3}$ pc) enlarging the star-to-star distance. The larger sonic length also implies a larger embedding 
%EJL
%cloud 
clump
surrounding the subject star such that the incoming flux will be more heavily attenuated. This heavy extinction through neutral ISM is the main culprit that completely removes the contribution of EUV flux in the solar neighborhood. We note that while \cite{fatada08} find a larger EUV flux in heavy extinction environments ${\sim}10^{3}$ photons s$^{-1}$ cm$^{-2}$, their results accord with our conclusion that EUV fluxes in such environs are too small to affect protoplanetary disk evolution: the \cite{fatada08} flux produces a mass loss rate (equation \ref{equation:mdot_euv}) ${\sim}10^{-12} M_{\odot}$ yr$^{-1}$, ${\sim}$two orders of magnitude below that catalyzed by the central star by EUV photons \citep[e.g.][]{gordulhol09} and four orders of magnitude below that of X-rays \citep[e.g.][]{picercowe19}.

We also observe that the lower velocity dispersion in the solar neighborhood yields a slower evolution in UV flux over time than in thick disk clouds. The modest increase in flux over time in the solar neighborhood is due to stellar evolution, as massive stars' radii expand slightly while still on the main sequence and increase their UV output. By $\sim$1 Myr, expansion of the cluster begins to dominate and the flux begins to drop. In the thick disk on the other hand, cluster expansion takes hold within ${\sim}10^{5}$ yr.

Within the primordial thick disk, we find both FUV and EUV flux to be larger in more massive clouds.
The radiation field on a subject star is primarily determined by the total stellar mass in the immediate vicinity of the target star (within $\sim$3 pc) which is necessarily larger in more massive clouds, both because of the larger total mass budget and because the more massive clouds sample the stellar IMF to higher masses.
This UV flux is further modulated by attenuation which is weaker in more massive thick disk clouds due to their higher velocity dispersion leading to lower $\rho_{\rm cl}$ (c.f.~equation \ref{equation:bulk_density}).
In all cases, FUV fluxes exceed EUV due to higher intrinsic stellar FUV fluxes and the lower opacity of the ISM to FUV radiation (equations \ref{equation:tau_euv} and \ref{equation:tau_fuv}). In particular, there is a steep drop in EUV flux at $\sim$1.5${\times}10^5$ yr in the $10^7 \ M_\odot$ cloud whereas the FUV flux continues to gradually decline. This is the time at which the distance between nearby massive stars and the subject star has grown beyond the Str{\"o}mgren radius: the EUV flux is exposed to neutral ISM, rapidly decreasing the ionization fraction and 
absorbing the radiation. The FUV flux slowly decreases due to cluster expansion, relatively unabated by the ISM. 

The middle panel of Figure \ref{figure:euvt} depicts an increase in $\Phi_{\rm UV}$ with larger $\epsilon_\star$ at a fixed cloud mass. For a given cloud mass, larger $\epsilon_\star$ implies a larger total stellar mass that populates the higher mass end of the stellar IMF, boosting the UV flux. For a fiducial thick disk efficiency of $\epsilon_{\star}=10\%$ (in blue), the initial UV flux is $\sim$5--6 orders of magnitude larger than for a solar neighborhood GMC (left panel). The FUV flux in our fiducial thick disk cloud (${\sim}10^{8}$ G0) exceeds the unattenuated flux estimated in the Central Molecular Zone (CMZ) by \cite{winkruche20} (${\sim}10^{7}$ G0). Our smallest FUV flux (${\sim}10^{6}$ G0, in our cluster of stellar mass $2{\times}10^{5} M_{\odot}$; left panel) is however comparable to their estimated fluxes (${\sim}10^{6}$ G0) in the densest CMZ cluster, Arches (${\sim}10^{4} M_{\odot}$; \cite{figkimmor99}), and in the most massive known cluster in the Local Group, Westerlund 1 (${\sim}10^{4-5} M_{\odot}$; see \cite{clanegcro05} and \cite{pormcmgie10}). We find no overlap however with the maximally attenuated fluxes of \cite{winkruche20}, which only reach up to ${\sim}10^{5}$ G0 \citep{winkruche20}. The largest stellar system we consider (of stellar mass $2.4{\times}10^{8} M_{\odot}$) produces an upper limit FUV flux of ${\sim}10^{9}$ G0. 
Our fluxes generally exceed those found by \cite{winkruche20} because we consider clusters up to ${\sim}$3 orders of magnitude more massive and because we account for stellar clustering, which both boost the UV field evaluated at a target star.

The right panel of Figure \ref{figure:euvt} illustrates the effect of mass segregation on the UV flux evolution, for a fixed cloud mass of $10^{8} M_{\odot}$ at $\epsilon_{\star}=10\%$. If stellar velocities are 
initially
mass-segregated to the extent that they reflect an equipartition-like distribution (illustrated in the blue curve, with $\sigma_{\star}\propto m^{-1/2}$), the primordial UV flux can persist for at least an order of magnitude longer than if there is no mass-dependence (in red). This effect arises because the radiation field is dominated by the most massive stars, so systematically decreasing their velocity dispersion keeps massive stars preferentially close to the subject star for longer.

\begin{figure}
\epsscale{1.2}
\plotone{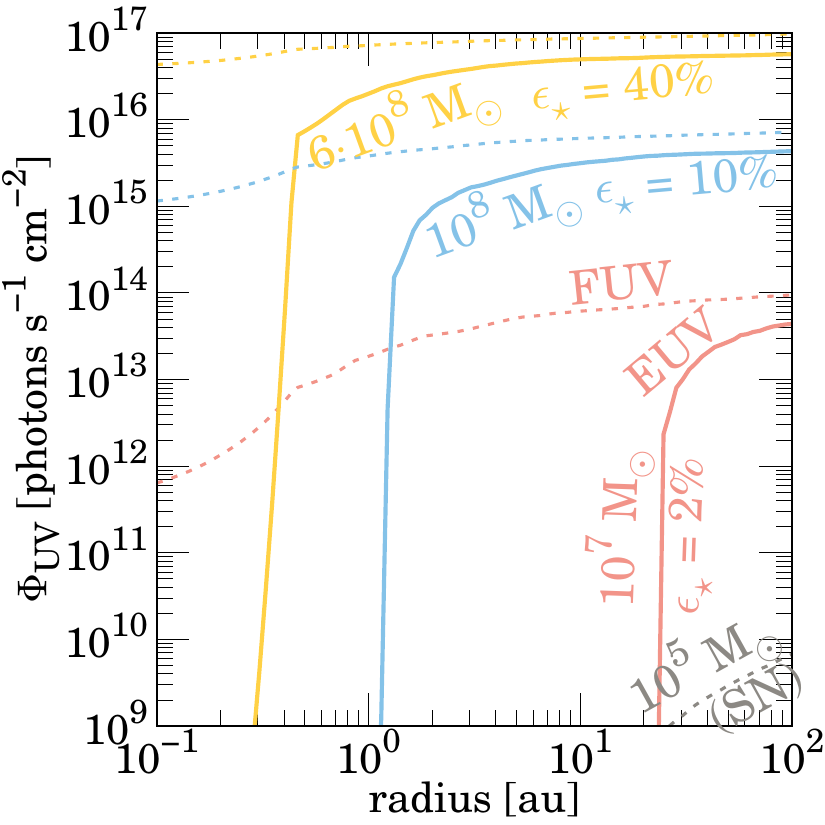}
\caption{Initial UV flux as a function of radial distance from the subject star. Cloud masses and star formation efficiencies follow the color scheme of Figure \ref{figure:euvt}. EUV fluxes are shown in solid lines, FUV in dashed. UV fluxes drop with decreasing radial distance due to attenuation through the protostellar envelope. Where EUV photons are completely extincted, FUV photons persist and can continue driving disk photoevaporation. EUV photons penetrate deeper through the protostellar envelope in more massive GMCs, illuminating a larger fraction of the disk.
\label{figure:euvr}}
\end{figure}

\begin{figure}
\epsscale{1.2}
\plotone{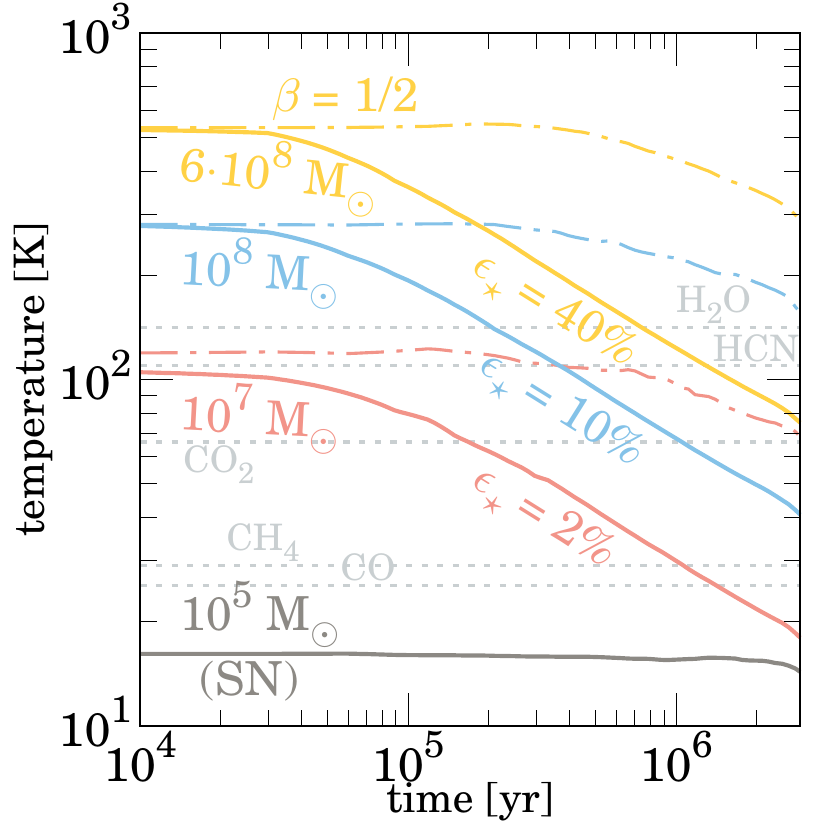}
\caption{Evolution tracks in temperature at the location of the subject star due to external bolometric radiation for $\beta = 0$ (solid lines) and $\beta = 1/2$ (dot-dashed lines). The color scheme follows that of Figure \ref{figure:euvr}. Similar to $\Phi_{\rm UV}$, the extended radiation in ``mass-segregated'' ($\beta = 1/2$) clouds lengthens the timescale over which the temperature remains raised compared to $\beta = 0$.
The bolometric temperature in primordial thick disk GMCs can exceed solar neighborhood conditions by more than an order of magnitude. Such high temperatures heat protoplanetary disks above common chemical species' condensation temperatures (plotted in gray dotted lines as mean temperatures from Table 2 of \cite{zhablaber15}) at early times before the cluster expands. Such high temperatures also promote protoplanetary disk dispersal through raising the sound speed which in turn raises kinematic viscosity. \label{figure:Tbolt}}
\end{figure}

\begin{figure}
\epsscale{1.2}
\plotone{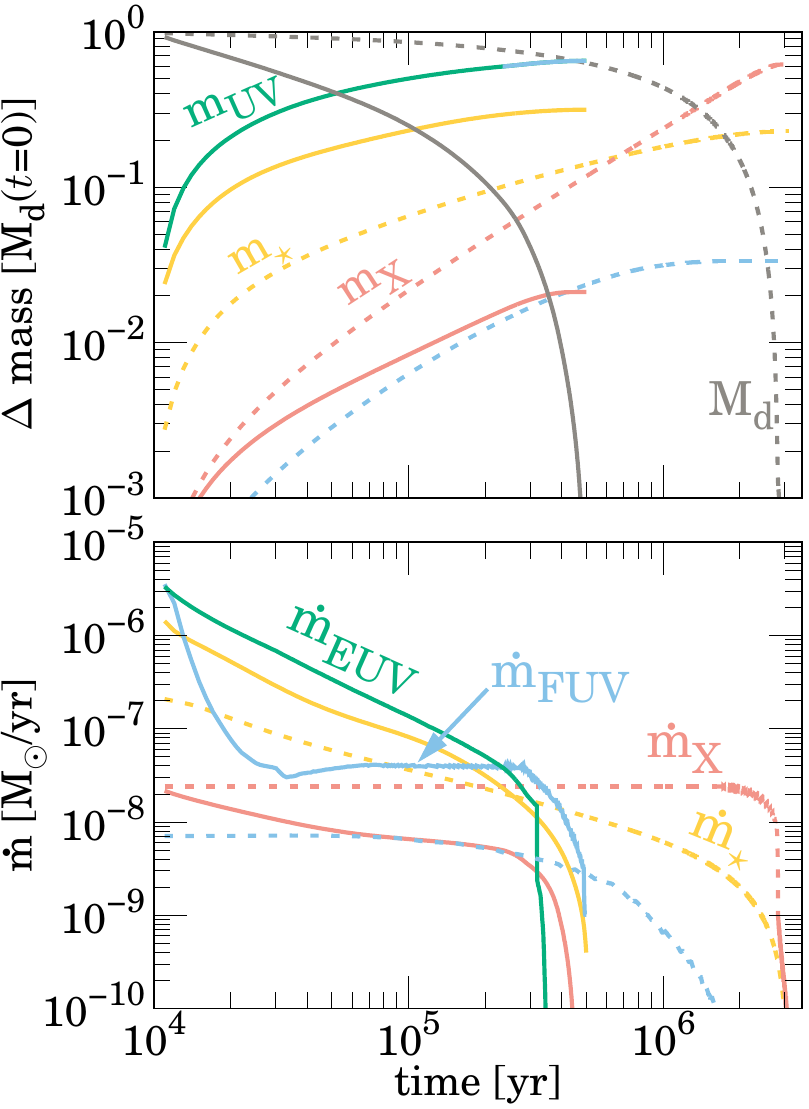}
\caption{Protoplanetary disk evolution in a solar neighborhood GMC ($10^{5} M_{\odot}, \ \epsilon_{\star}=2\%$, dashed lines), and in a GMC in the primordial thick disk ($10^{8} M_{\odot}, \ \epsilon_{\star}=10\%$, solid lines). Top: total mass lost through stellar accretion (in yellow; $m_{\star}$), internal X-ray evaporation (in red; $m_{\rm X}$), and external FUV and EUV evaporation (in blue and green, respectively; $m_{\rm UV}$) through time. Total disk mass is shown in gray ($\mathrm{M}_{\rm d}$) as a reference. Middle: the mass loss rates for each process. Internal evaporation sets the disk lifetime to ${\sim}$3 Myr in the solar neighborhood, while external EUV and FUV-driven evaporation truncate the disk lifetime to ${\sim}$0.5 Myr in the primordial thick disk.
\label{figure:mass_compare}}
\end{figure} 

In Figure \ref{figure:euvr} we display UV flux versus radius from the subject star, at time $t=10^{4}$ yr. For each curve, the UV flux declines toward smaller radial distances as it is absorbed through the protostellar envelope. The EUV flux is completely extincted at a radius inside the disk outer edge, a location we refer to as the ``EUV extinction radius". In the most massive stellar environments ($6{\times}10^{8} M_{\odot},  \ \epsilon_{\star}=40\%$), almost the entire protoplanetary disk is illuminated by EUV photons. In the least massive environment ($10^{7} M_{\odot}, \ \epsilon_{\star}=2\%$), only the outer reaches of the disk beyond $\gtrsim 30$ au are illuminated by the EUV flux. Disks in the environments shown in Figure \ref{figure:euvr} can undergo EUV-driven evaporation from radii beyond the EUV extinction radius. Inside the EUV extinction radius, the disk remains illuminated by FUV photons so the mass loss can only be FUV-driven. Over time as the cluster expands, the $\Phi_{\rm UV}$ curves will decrease in magnitude and the EUV extinction front will move to greater distances. In the solar neighborhood on the other hand (bottom right corner of Figure \ref{figure:euvr}), no EUV photons are present and FUV photons impinge only on the outermost regions of the disk $\sim$100 au.

Figure \ref{figure:Tbolt} illustrates evolution tracks in the temperature at the location of the subject star due to external bolometric radiation. 
For our fiducial thick disk cloud (blue curves), we find an initial background temperature $\sim$300 K, $\sim$20$\times$ larger than the $\sim$17 K solar neighborhood. This flux ($F_{\rm cl}$ in equation \ref{equation:disk_temp}) will dominate other energy sources down to distances $\sim$5 au from the host star, inside which viscous dissipation contributes the lion's share. The overall rise in the disk temperature has an important consequence on the disk dissipation timescale. The kinematic viscosity $\alpha c_s H \propto T$ so that the viscosity across much of a protoplanetary disk in the $10^{8} M_{\odot}, \ \epsilon_{\star}=10\%$ environment can exceed that in the solar neighborhood by at least an order of magnitude. Under viscous accretion, such high viscosities promote more rapid disk dispersal and increase the rate at which the disk can supply mass to any loss channels (e.g. evaporative winds). 
Over time, the bolometric heating subsides as the cluster expands, although in mass-segregated clouds, the timescale over which the disks experience intense bolometric heating lengthens owing to massive stars lingering around the target star. 

The increase in background temperature also has important implications for disk chemical evolution. At early times before the clusters expand, temperatures in our thick disk clouds exceed the condensation temperatures of major volatile species \citep[see also][for further consideration of this effect in the vicinity of a single massive star]{haw21}. Since mass-segregated environments remain hotter for longer timescales than non-segregated environments, even water can be sublimated in the largest segregated clouds for $\gtrsim$Myr timescales. 

We next explore how the intense radiation environments illustrated in Figures \ref{figure:euvt}-\ref{figure:Tbolt} impact the evolution of protoplanetary disk gas surface density.

\begin{figure}
\epsscale{1.2}
\plotone{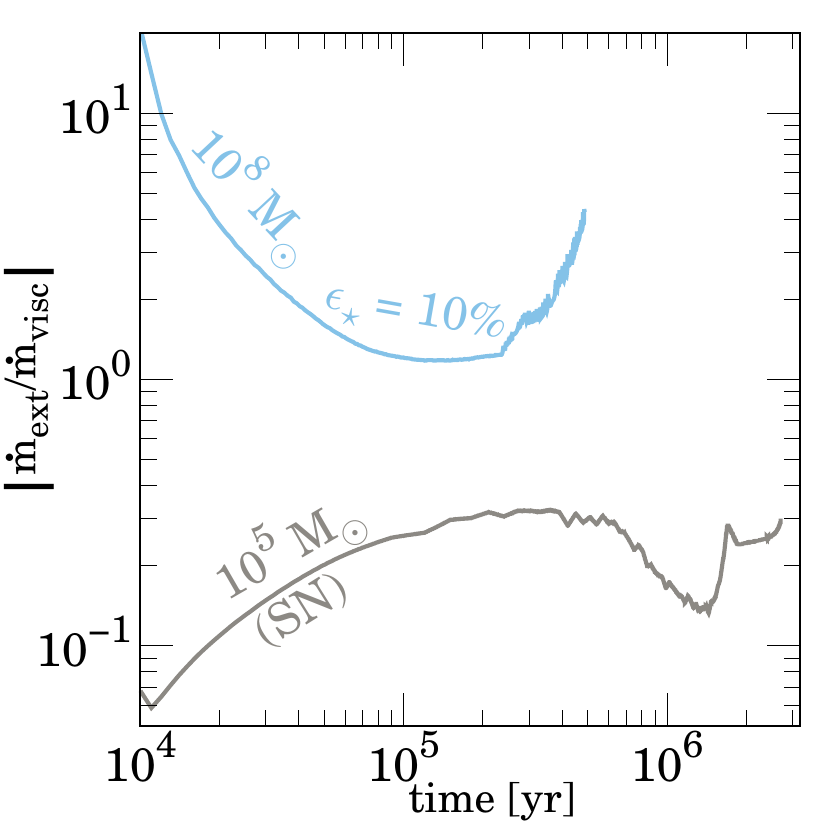}
\caption{The ratio of external evaporation rate $|\dot{m}_{\rm ext}|$ to the viscous mass flow rate $|\dot{m}_{\rm visc}|$ near the disk outer radius $R_{\rm disk}$, during protoplanetary disk evolution in our solar neighborhood GMC (gray) and fiducial thick disk GMC (blue). The viscous mass flow rate is calculated by averaging equation \ref{equation:mdot_disk} over 150 cells interior to the innermost cell that externally evaporates (which is near $R_{\rm disk}$). In the solar neighborhood, viscous spreading replenishes mass lost to external evaporation, and mass decretes through the simulation grid's outer boundary (we use an outflow boundary condition). In the primordial thick disk, external evaporation outpaces viscous spreading, quickly eroding $R_{\rm disk}$.
\label{figure:mdcomp}}
\end{figure} 

\subsection{Protoplanetary Disk Evolution}\label{subsec:disk_evo}

In Figure \ref{figure:mass_compare}, we illustrate the evolution of total mass lost, evaporation rate, and stellar accretion rate in our fiducial Galactic thick disk conditions compared to that in the solar neighborhood.

\begin{figure*}
\centering
\includegraphics[width=\textwidth]{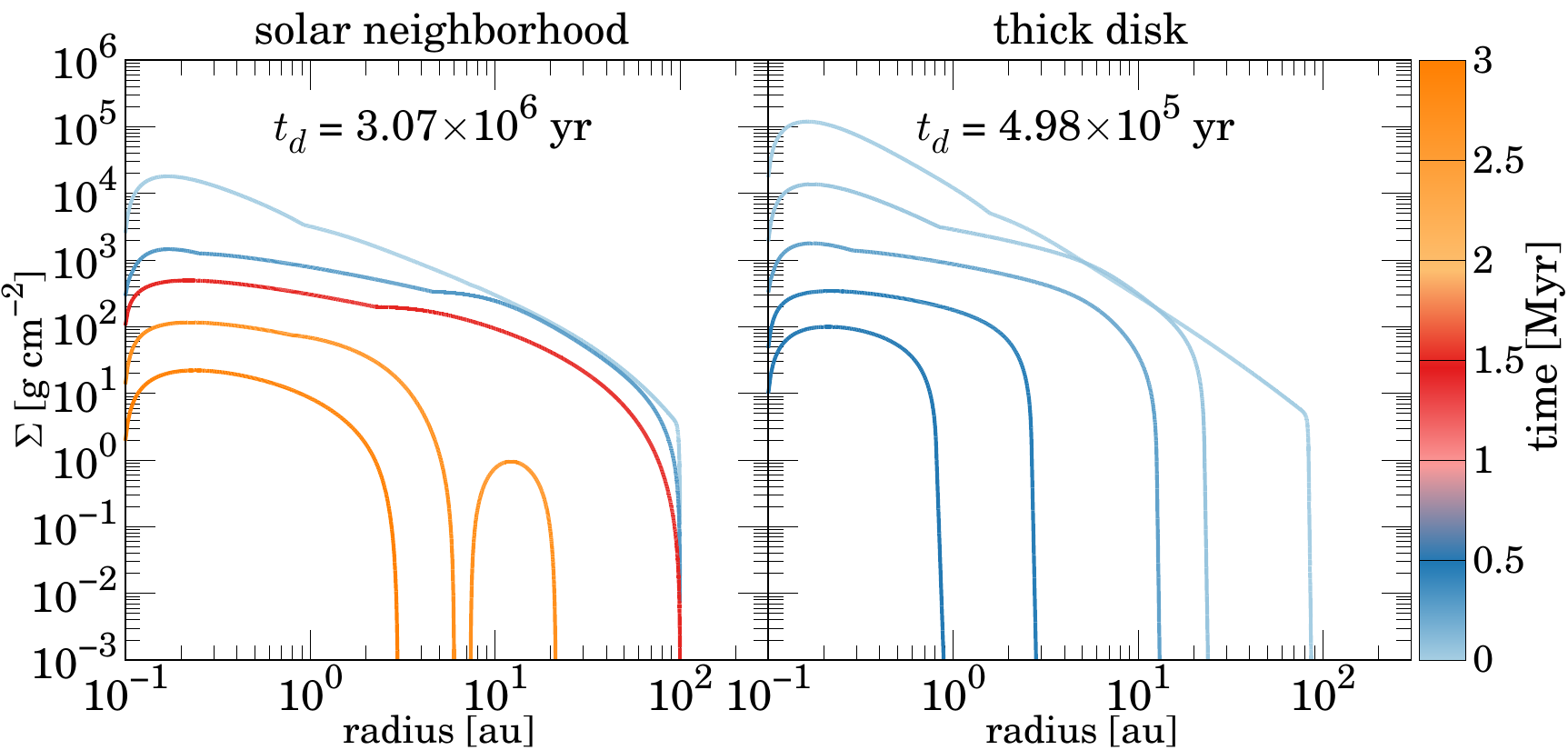}
\caption{Gas surface density evolution of the two protoplanetary disks shown in Figure \ref{figure:mass_compare}. Different colors correspond to different epochs according to the color bar to the right of the plot. The solar neighborhood protoplanetary disk passively spreads for ${\sim}2$ Myr before being dispersed by internal X-ray evaporation by ${\sim}3$ Myr ($t_{\rm d}$ labels the disk lifetime). In the primordial thick disk, ambient UV photons drive a rapid wind that empties out the disk in ${\sim}0.5$ Myr. [An animated version of this figure is available in the HTML version of this paper, as well as at \url{https://thallatt.github.io/thick_disk}] \label{figure:sn_compare}}
\end{figure*} 

The solar neighborhood disk undergoes two evolutionary epochs: a long term phase of passive viscous spreading ($t$=0--2 Myr), followed by a late stage dominated by internal, X-ray evaporation which ultimately sets the disk lifetime (when the disk mass falls below $10^{-5}M_{\odot}$, at 3 Myr). External evaporation hardly affects the disk. The initial FUV flux, as illustrated in the left panel of Figure \ref{figure:euvt}, corresponds to ${\sim}$70 G0, translating to an external evaporation rate that remains lower than any other mass loss channel at all times. 

The protoplanetary disk in the cosmic noon environment suffers elevated external evaporation, driven by the significantly larger UV flux from the surrounding cluster than in the solar neighborhood. The initial EUV flux is sufficiently high (see Figure \ref{figure:euvt}) that FUV-heated gas in the disk surface cannot pass through a sonic point (i.e. $x_{\rm PDR}{\ll}1$), so that the evaporation is EUV-driven for the bulk of the evolution. The FUV mass loss rate also declines much more quickly 
%EJL
%with decreasing radius 
%
than EUV initially, likely due to the cooler temperature of the FUV-heated outflow. As 
disk 
radii erode inside the FUV Bondi radius $G M_{\star} / c^{2}_{\rm s, FUV}{\sim}100$ au, the gas kinetic energy falls below the gravitational
potential energy, helping confine the outflowing gas. The kinetic energy of EUV-heated gas dominates over gravity beyond ${\gtrsim}$9 au, which combined with increasing density at the base of the EUV outflow at smaller radii due to ionization-recombination balance (in which the
density at the outflow’s ionization front $n_{\rm IF}{\propto}R_{\rm disk}^{-1/2}$; \cite{johholbal98}), helps produce a weaker drop in mass loss rate with decreasing radius. Once the disk has eroded to the point that the protostellar envelope extincts all EUV flux, FUV mass loss takes over until the disk is destroyed.

Dispersal of the protoplanetary disk in the thick disk GMC is aided by its higher stellar accretion rate relative to the solar neighborhood. The accretion rate is elevated partly due to the difference in initial condition we adopt, which loads more mass in the inner disk (equation \ref{equation:IC}). \footnote{Re-running the solar neighborhood simulation with the initial conditions in $\gamma$ and $d\ln H/d\ln R$ used in the thick disk yields a shorter lifetime ${\sim}2.6$ Myr, due to the increased stellar accretion rate.} 
%EJL
Additionally, external evaporation promotes viscous evolution by truncating the disk and thereby shortening the global viscous timescale ($t_\nu {\sim}R^2_{\rm disk}/\nu$),
%EJL
which is further shortened by elevated $\nu$ from background heating (see Figure \ref{figure:Tbolt}).

\begin{figure*}
\centering
\includegraphics[width=1\textwidth]{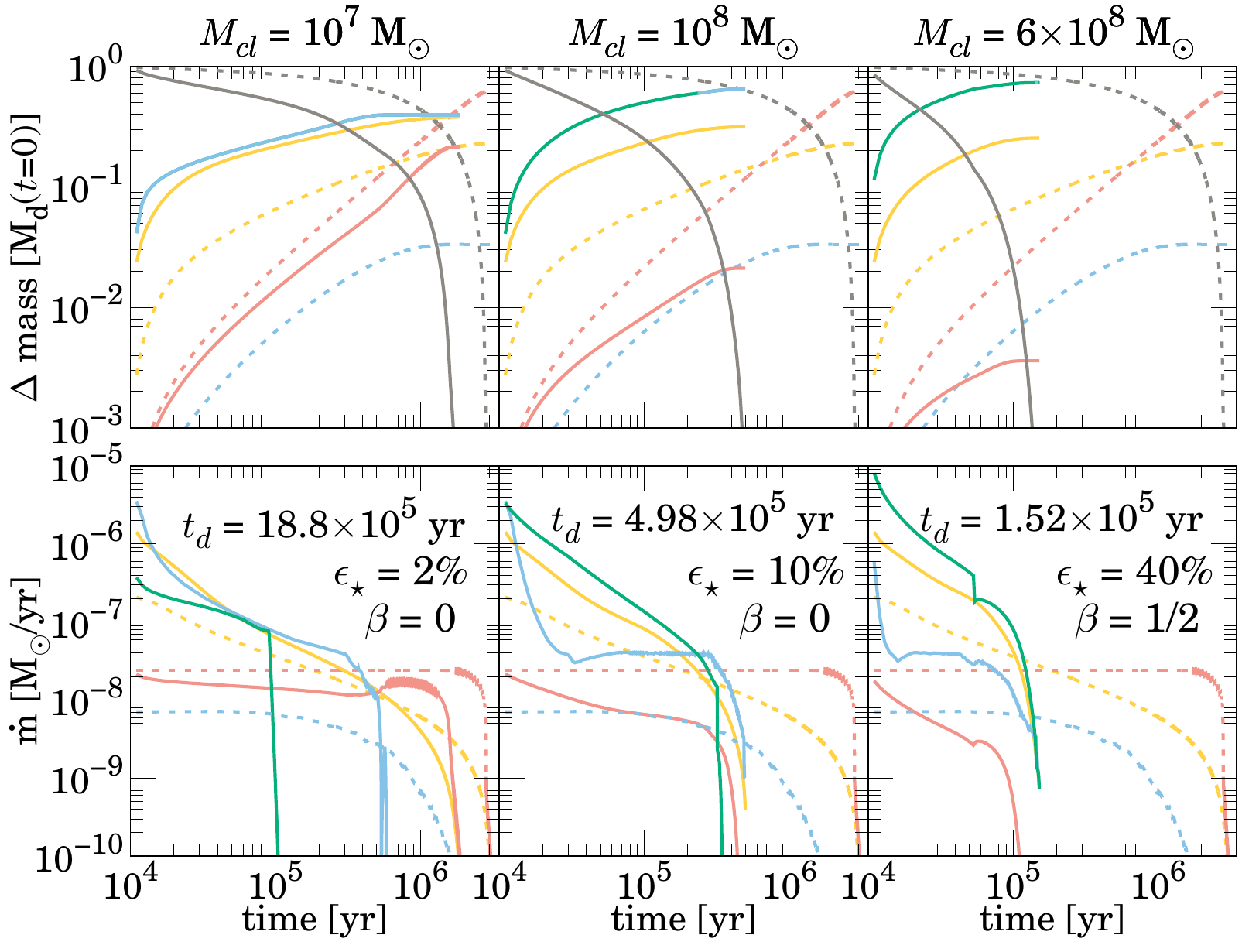}
\caption{Protoplanetary disk evolution in the least and most extreme GMCs considered in this work: $10^{7}M_{\odot}, \epsilon_{\star}=2\%$ with non-segregated stellar dynamics (left panel), our fiducial $10^{8} M_{\odot}, \epsilon_{\star}=10\%$ (middle), and $6\times 10^{8}\ M_{\odot},\epsilon_{\star}=40\%$, with mass-segregated dynamics (right). The color and line type scheme adheres to Figure \ref{figure:mass_compare}. Disks in the least extreme environments dissipate after $\sim$2 Myr under the combined effect of moderate FUV evaporation, stellar accretion, and internal X-ray evaporation. Once FUV evaporation abates by $\sim$0.5 Myr, the disk viscously spreads before internally sourced X-rays set its lifetime. Disks in the most extreme environments suffer intense external EUV evaporation, which truncates disk lifetimes to just ${\sim}0.1$ Myr. \label{figure:mdot_paramspace}}
\end{figure*}

Figure \ref{figure:mdcomp} illustrates why $R_{\rm disk}$ is truncated rapidly in the primordial thick disk. In the solar neighborhood, the external mass loss rate is always less than the outward viscous mass flow rate. The disk radius is consequently never affected by external evaporation, remaining fixed at the 100 au outer boundary of our computational domain until internal evaporation disperses the disk (enlarging our domain would allow the disk to expand outwards during the viscous spreading epoch). In our thick disk GMC, the external mass loss rate exceeds the outward mass supply rate by more than an order of magnitude initially. Unable to replenish the mass that is lost, $R_{\rm disk}$ erodes inwards until it stalls when evaporation is balanced by viscous diffusion at ${\sim}10^{5}$ yr. Evaporation once again outpaces the disk decretion rate once FUV drives the mass loss (see the uptick at ${\sim}3{\times}10^{5}$ yr), eating away at $R_{\rm disk}$ until it is destroyed.

\begin{figure*}
\centering
\includegraphics[width=1\textwidth]{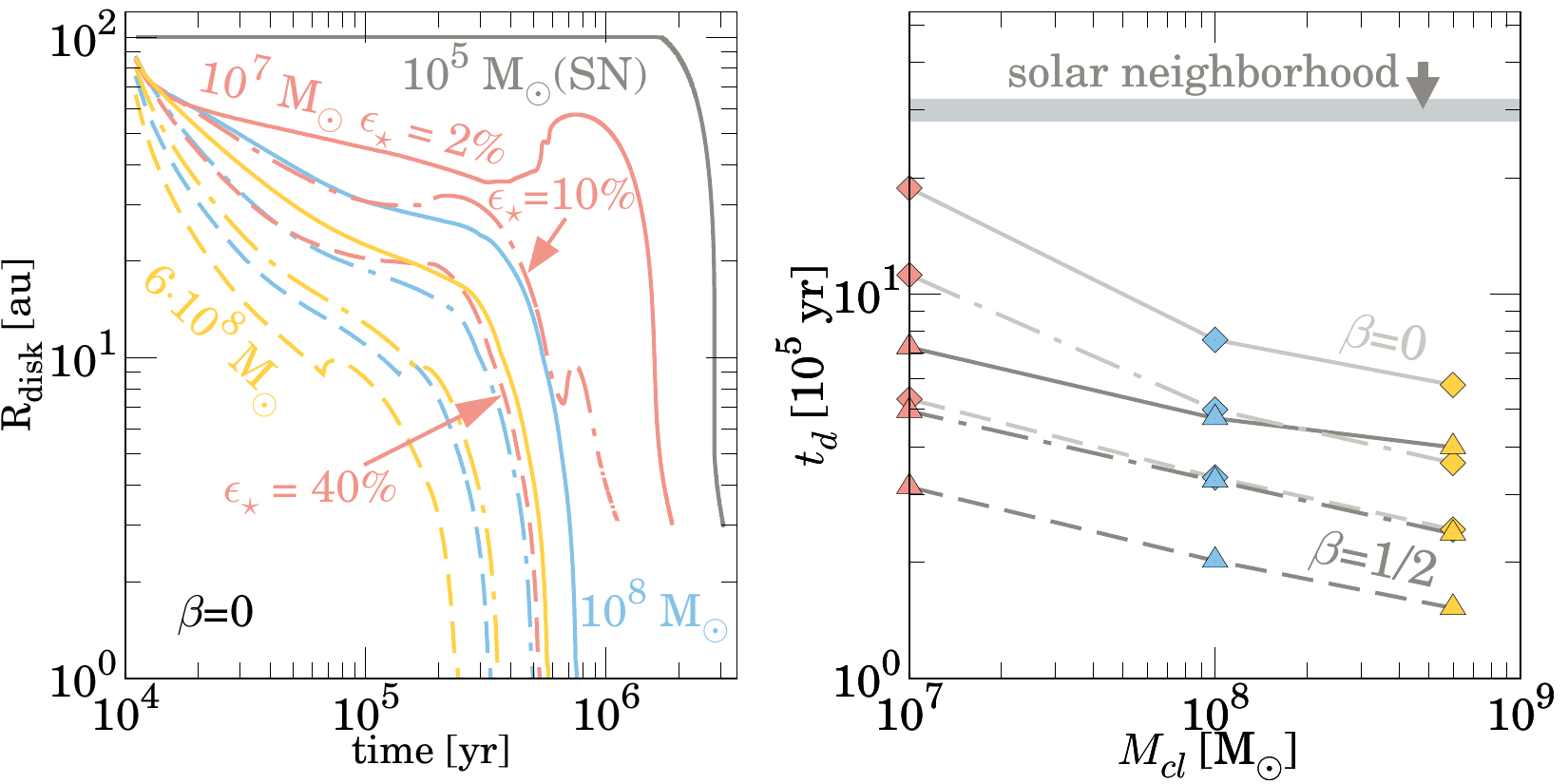}
\caption{Left: protoplanetary disk radius (defined as the outermost cell where $\Sigma{>}10^{-4}$ g cm$^{-2}$) evolution in each non-mass segregated environment we consider. The mass of each GMC follows the convention of Figure \ref{figure:euvt}. Environments with $\epsilon_{\star}=2,10,40\%$ are depicted in solid, dot-dashed, and dashed lines, respectively. Protoplanetary disk radii erode inwards more rapidly as external evaporation increasingly overwhelms viscous diffusion in more extreme environments (see also Figure \ref{figure:mdcomp}). Right: protoplanetary disk lifetime $t_{d}$ versus GMC mass. Data points are color-coded following the left panel. Environments with $\epsilon_{\star}=2,10,40\%$ are connected via gray lines that follow the same line convention as the left panel. Environments with non-mass segregated ($\beta=0$) and mass segregated ($\beta=1/2$) stellar dynamics are shown as diamonds and triangles joined by light and dark gray lines, respectively. Protoplanetary disks are destroyed more rapidly in environments with higher $M_{\rm cl}$ and $\epsilon_{\star}$, and in mass-segregated stellar populations. Elevated $M_{\rm cl}$ and $\epsilon_{\star}$ during the era of thick disk emergence 
%EJL
can
conspire to shorten protoplanetary disk lifetimes by up to $\sim$an order of magnitude.
\label{figure:rdisk_tdisk}}
\end{figure*} 

Figure \ref{figure:sn_compare} illustrates the gas surface density 
%EJL
profile
through time for our protoplanetary disks in the solar neighborhood and fiducial thick disk GMCs. Differences in initial condition between solar neighborhood and thick disk $\Sigma(r)$ are evident in the snapshots at $t=10^{4}$ yr inside ${\lesssim}1$ au. Figure \ref{figure:sn_compare} shows that for the first ${\sim}$2 Myr of evolution in the solar neighborhood, the disk surface density decreases almost uniformly under viscous diffusion. Internal evaporation only visibly affects the disk's surface density once the rate of viscous mass supply into a given cell drops below the local internal evaporation rate. We find that this threshold is first crossed at $R_{\rm disk}$, after ${\sim}$2 Myr. The disk outer edge proceeds to be eaten down to ${\sim}$20 au before a gap in surface density ($\Sigma\rightarrow0$) is cleared at $\sim$7 au where $\dot{\Sigma}(r)$ from internal evaporation peaks (see the snapshot at $t=0.9\times t_{\rm d}$). Having lost the supply of gas to replenish what is accreted, the inner disk drains out onto the star over its local viscous time, while the remnants of the outer disk evaporate away \cite[see e.g.][]{clagensot01}. We note that the destruction of the outer disk we find during the X-ray-driven dispersal phase differs from previous work that reported inside-out disk dispersal with no outer disk destruction \citep[e.g.][]{oweerccla10}. We find that, during the X-ray dominated phase, whether disk radii erode inwards or disks disperse inside-out without radius truncation depends on the X-ray evaporation mass loss profile $\dot{\Sigma}_{\rm int}$ we use. Employing $\dot{\Sigma}_{\rm int}$ from Appendix B of \cite{oweclaerc12} yields inside-out dispersal, in agreement with previous studies \citep[e.g.][]{oweerccla10}, whereas our fiducial $\dot{\Sigma}_{\rm int}$ from \cite{picercesp21} produces outside-in erosion of $R_{\rm disk}$. The wind mass loss profile of \cite{picercesp21} is more efficient at removing material at large radii, and less strongly peaked inside $R{\lesssim}$a few au, than \cite{oweclaerc12} (the former possibly a result of their numerical scheme resolving lower ionization parameters, allowing more accurate modelling of disk outer radii; \cite{picercesp21}), which initially depletes the outer disk faster than the inner disk. The disk evolution shown in the right panel of Figure \ref{figure:sn_compare} highlights the rapid outside-in destruction caused by external evaporation at cosmic noon (for a comprehensive overview of disk evolution under both internal and external evaporation in less extreme environments, see e.g. \cite{colhaw22} and \cite{colmrohaw24}).

Our finding that disks around solar mass stars in the solar neighborhood undergo negligible external evaporation may be at odds with the decrease in median disk dust mass with ambient FUV flux measured in Orion by \cite{vanhac23}. This discrepancy could arise due to a preponderance of K and M-type stars in the survey \citep{vanhacvan22} whose shallower potential wells admit enhanced photoevaporation rates relative to solar type stars \citep{hawclarah18}. The decrease in dust mass observed by \cite{vanhac23} around low mass stars can be reconciled with the observation that M stars appear to host longer lived disks than solar type stars (in massive clusters \citep{fanboekin12} as well as nearby low mass clusters \citep{pfadehmic22}) due to the fact that the dust mass inferred by \cite{vanhac23} is measured from millimeter luminosity originating from millimeter sized grains in the outer disk, whereas the disk fractions employed by \cite{pfadehmic22} and \cite{fanboekin12} use emission from near to mid infrared originating from micron sized grains in the inner disk. In this picture, disks around low mass stars in the SODA survey may indeed suffer dust mass loss through evaporation from the outer disk. Such dust advection must occur early on before grains grow too large to entrain in the gas \citep[e.g.][]{hawfaccla18} and before drifting inward after ${\sim} 10^{5}$ yr to radii where evaporation is less efficient \citep[e.g.][]{selboocla20}. After the early loss of dust, evaporation may shut down as low mass stars move away from massive stars through dynamical mass segregation. Micron sized grains could then persist in the remnant inner disk that slowly accretes \cite[if they are replenished by fragmentation of larger grains; e.g.][]{birklaerc12}.

We next study the effect of varying the star formation efficiency, GMC mass, and degree of mass segregation. Figure \ref{figure:mdot_paramspace} brackets our parameter space, illustrating protoplanetary disk evolution in the least and most extreme thick disk environments. Figure \ref{figure:mdot_paramspace} demonstrates that in the least extreme environments, protoplanetary disks dissipate under the combined effect of FUV evaporation and stellar accretion, both of which consume equal amounts of the disk. The initial FUV flux is $\sim$4 orders of magnitude larger than in the solar neighborhood (see Figure \ref{figure:euvt}), translating to an increase in $\dot{m}_{\rm FUV}$ over the solar neighborhood by up to $\sim$3 orders of magnitude. The disk undergoes FUV mass loss until cluster expansion aided by attenuation through the protostellar envelope extinct all FUV photons (see Figure \ref{figure:euvt}). Once FUV evaporation tapers off, the disk is free to viscously expand from $\sim$30 to 60 au, increasing the surface area of the disk capable of undergoing internal X-ray evaporation. The X-ray evaporation rate increases, ultimately setting the disk lifetime. 

At the other end of parameter space, Figure \ref{figure:mdot_paramspace} highlights that disks in the most extreme environments are rapidly destroyed, solely by EUV-driven evaporation. Since the EUV mass loss rate increases with the number of recombinations in the flow (\cite{johholbal98}; see also \cite{holjohliz94}), while the FUV mass loss rate plateaus above fluxes $F_{\rm FUV} \gtrsim 10^{4-5}$ G0 \citep[][]{stohol99,winhaw22}, $\dot{m}_{\rm EUV}$ exceeds $\dot{m}_{\rm FUV}$ by at least an order of magnitude for almost the entire disk lifetime. The disk is extremely quickly eroded down to the EUV Bondi radius $\sim$9 au, at which time the mass loss rate abruptly drops as the two EUV evaporation models of \cite{johholbal98} and \cite{owelin23} are discontinuous at ${\sim}$9 au. The discontinuity likely stems from the PDR model of \cite{owelin23}, which assumes that the FUV-heated gas is extremely strongly bound ($R_{\rm disk}{\ll}$the FUV Bondi radius ${\sim}$100 au), and therefore may help confine the mass loss. We have verified that differences in EUV mass loss rate produce a difference in disk lifetime of at most ${\sim}10^{5}$ yr. We also observe a slight uptick in the disk radius and X-ray evaporation rate at this time due to the drop in $\dot{m}_{\rm EUV}$. Ultimately, EUV-driven evaporation evacuates ${\sim}80\%$ of the disk mass, and the disk is destroyed ${\sim}20\times $ faster than in the solar neighborhood. 

Our results are summarized in Figure \ref{figure:rdisk_tdisk}, which showcases disk radius evolution and disk lifetimes across the range of environments we consider. The competition between viscous spreading and mass loss displayed in Figure \ref{figure:mdcomp} reflects $\dot{R}_{\rm disk}$ for the fiducial thick disk curve in the left panel of Figure \ref{figure:rdisk_tdisk}. In the least extreme thick disk cloud (also plotted in Figure \ref{figure:mdot_paramspace}), $\dot{R}_{\rm disk}$ evolves as the dominant disk evolution process changes through time. The disk first gradually erodes under FUV evaporation, before viscously expanding once the FUV radiation field dwindles to zero. Internal X-ray evaporation then ultimately drives $R_{\rm disk}\rightarrow 0$ once the viscous mass flux at $R_{\rm disk}$ falls below the internal mass loss rate. In the most extreme environments, viscous spreading is almost never competitive with EUV evaporation (save for the brief moment when $\dot{m}_{\rm EUV}$ suddenly drops as $R_{\rm disk}$ erodes inside the EUV Bondi radius), leading to rapid, almost continuous $R_{\rm disk}$ erosion. 
%EJL
%We find that varying each environmental parameter ($M_{\rm cl}, \epsilon_{\star}, \beta$) in isolation shortens disk lifetimes by less than an order of magnitude. However, $M_{\rm cl}\propto \sigma^{4}_{\rm gal}$ is concomitant with $\epsilon_{\star}\propto \Sigma_{\rm gal}$ \citep[e.g.][]{gruhopfau18} in the highly turbulent and dense conditions of cosmic noon. These parameters can conspire to shorten disk lifetimes by $\sim$an order of magnitude compared to the solar neighborhood. 
While varying each environmental parameter ($M_{\rm cl}, \epsilon_{\star}, \beta$) individually shortens disk lifetimes by factors of just order unity, 
towards higher $M_{\rm cl} \gtrsim 10^8 M_\odot$ and highest $\epsilon_\star \sim$40\%, the disk lifetimes can shorten by up to an order of magnitude.

The disk evolution simulations we have presented thus far used identical initial disk masses ($0.1 M_{\odot}$) and radii (100 au). We next consider how variations in disk initial radii or masses may affect our disk lifetimes. Larger initial disk radii in our cosmic noon clouds will be eroded down to our fiducial radius of 100 au extremely quickly under boosted external evaporation rates ($\dot{m}_{\rm EUV}{\propto}R^{3/2}_{\rm disk}$), erasing the difference in initial condition. More compact disks would still suffer intense evaporation (see e.g. Figure \ref{figure:euvr}) while also accreting more rapidly due to their higher surface density (assuming fixed mass), likely still emptying out over ${\sim}$a few ${\times}10^{5}$ yr. In the solar neighborhood, the increase in FUV evaporation rate around more extended disks is unlikely to shorten disk lifetimes because the bulk of the disk mass is still contained inside ${\sim}$100 au where X-ray evaporation sets the dispersal timescale. We do expect our quoted lifetimes to lengthen sub-linearly with increasing disk mass, as heightened viscous accretion around more massive disks offsets the increase in evaporation timescale.

\section{Discussion}\label{sec:discussion}

\subsection{Disk Irradiation in Optically Thick Clouds}\label{sec:opt_thick}

Thus far, we have considered GMCs with canonical ISM dust abundance $\eta_{\rm d}=1\%$. Such clouds are optically thin to UV photons. Environments that are optically thick however (i.e. higher $\eta_{\rm d}$) may change disk irradiation conditions and evolution in two ways. First, clouds that are sufficiently optically thick to extinct UV photons over short distances may shut down external disk evaporation. Second, in GMCs that are globally optically thick to far-infrared (FIR) radiation re-emitted by dust, the temperature inside the GMC can exceed the bolometric temperature of the cluster following radiative diffusion \citep[e.g.][]{mur09,tho13}. If the FIR optical depth is sufficiently high, disk evolution in this case will no longer be dominated by external evaporation but by strong viscous dissipation (since $\nu{\propto}T$). The goal of this section is to estimate the dust abundance $\eta_{\rm d}$ above which disks cannot externally evaporate but may nonetheless suffer intense FIR heating, across the range of cloud bulk densities we consider in this paper. We will show that for the entirety of parameter space, disks evaporate and/or are heated by strong background FIR radiation.

We begin our analysis by computing the maximum dust abundance $\eta_{\rm d, UV}$ below which UV photons can still reach the subject star. To do so, we consider a random star in a ``clustered" environment (with correlation function $\propto r^{-2}$), separated from the nearest massive star by a typical distance $d_{\rm min}{\sim} 10^{-2}$ pc. We set the optical depth to EUV and FUV photons (equations \ref{equation:tau_euv} and \ref{equation:tau_fuv}, respectively) to unity over a distance of $d_{\rm min}$, and solve for the requisite dust fraction at a given cloud bulk density $\rho_{\rm cl}$ (set by $\sigma_{\rm gal}$ according to equation \ref{equation:bulk_density}). Since the ISM UV opacity is dominated by dust, setting the optical depth $\tau_{\rm UV}{\sim}\kappa_{\rm d, UV}\rho_{\rm cl} \eta_{\rm d} d_{\rm min}$ (taking just the dust component of density  $\rho_{\rm cl}\eta_{\rm d}$, since our $\kappa_{\rm d, UV}$ is defined per gram of dust; \cite{kuiyormig20}) to unity produces,

\begin{equation}\label{equation:eta_uv}
\eta_{\rm d, UV}=0.8\bigg(\frac{\kappa_{\rm d, UV}}{4{\times}10^{4} \ \mathrm{cm^{2} \ g^{-1}}}\bigg)^{-1}\bigg(\frac{\rho_{\rm cl}}{10^{-21} \ \mathrm{g \ cm^{-3}}}\bigg)^{-1}.
\end{equation}

We next calculate the minimum dust abundance $\eta_{\rm d, FIR}$ above which the GMC becomes globally optically thick to FIR. The FIR optical depth $\tau_{\rm cl, FIR} {\sim}\kappa_{\rm R}(\eta_{\rm d})\Sigma_{\rm cl}$ determines whether the radiation field at the subject star contains contributions from photons that are trapped and diffusing randomly through the GMC. We assume the FIR opacity scales linearly with dust fraction \citep[][]{tho13} via $\kappa_{\rm R}(\eta_{\rm d})=\kappa_{\rm R, 0} (\eta_{\rm d}/1\%)$ cm$^{2}$ g$^{-1}$, with $\kappa_{\rm R, 0}=5$ cm$^{2}$ g$^{-1}$ the normalization at $\eta_{\rm d}=1\%$ given by the \cite{semhenhel03} tables (appropriate for $T{>}150$ K; see also discussion surrounding equation \ref{equation:disk_temp}), and where $\kappa_{\rm R}$ is defined per gram of dust and gas mixture (as opposed to our UV opacities, which are given per gram of dust). Setting $\tau_{\rm cl, FIR}=1$ yields,

\begin{equation}\label{equation:eta_fir}
\begin{split}
\eta_{\rm d, FIR} {\sim}3{\times}10^{-3}\bigg(\frac{\kappa_{\rm R, 0}}{5 \ \mathrm{cm^{2} \ g^{-1}}}\bigg)^{-1} {\times} \\ & \hspace*{-4cm} \bigg(\frac{\Sigma_{\rm gal}}{1.7{\times}10^{3} M_{\odot} \ \mathrm{pc^{-2}}}\bigg)^{-1}\bigg(\frac{\phi_{\Sigma}}{2}\bigg)^{-1}. \\
\end{split}
\end{equation}

\begin{figure}
\epsscale{1.2}
\plotone{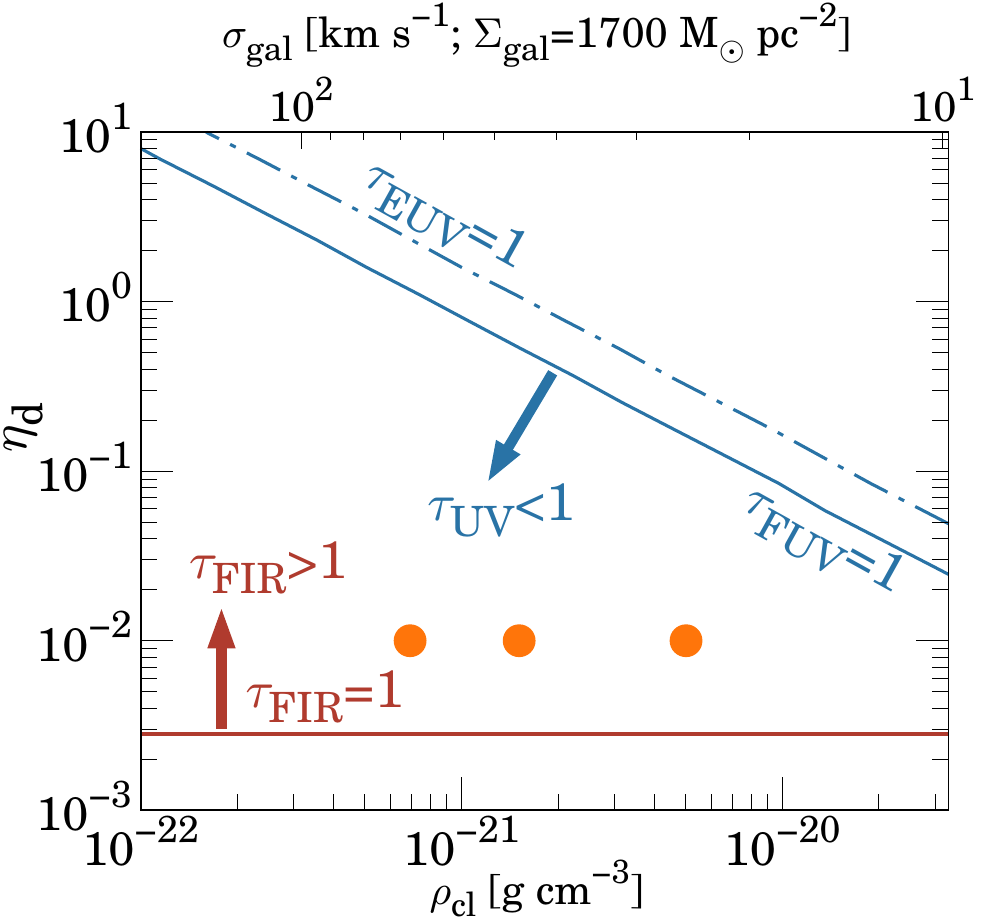}
\caption{Parameter space in ISM dust-to-gas mass fraction $\eta_{\rm d}$, GMC bulk density $\rho_{\rm cl}$, and Galactic disk velocity dispersion $\sigma_{\rm gal}$ (related to $\rho_{\rm cl}$ via equation \ref{equation:bulk_density}) over which GMCs are optically thick to far-infrared radiation (FIR) (and therefore exceed the cluster's effective temperature by $\tau^{1/4}_{\rm FIR}$, the region above the red line), and optically thin to FUV/EUV photons over typical intra-cluster stellar distances (below the blue solid and dot-dashed lines, respectively). Our thick disk GMCs are displayed in orange dots. Where $\tau_{\rm UV}<1$, protoplanetary disks can suffer external evaporation. Across the vast majority of parameter space, disks externally evaporate and/or are heated above the cluster's bolometric temperature. \label{figure:tau}}
\end{figure}

Figure \ref{figure:tau} compares $\eta_{\rm d, UV}$ to $\eta_{\rm d, FIR}$ across the range of cloud densities we consider. Clouds at cosmic noon are optically thick to FIR above $\eta_{\rm d}{\gtrsim}3{\times}10^{-3}$ and are therefore heated above the cluster's bolometric temperature by a factor ${\sim}\tau_{\rm cl, FIR}^{1/4}$ \citep[e.g.][]{mur09} (we 
%EJL
%chose 
choose
a lower limit of $\eta_{\rm d}=10^{-3}$ in Figure \ref{figure:tau} to reflect the lower limit of the observed distribution of thick disk stellar metallicity [Fe/H]${\sim}-$1; \cite{redlamall06}). To contextualize this estimate with respect to solar neighborhood clouds, we adopt a solar neighborhood GMC temperature $T{\sim}15$ K \citep{heydam15}, opacity $\kappa_{\rm R}(T,\eta_{\rm d})\propto T^{2}\eta_{\rm d} \sim 5\times 10^{-2} (\eta_{\rm d}/1\%) \ \rm cm^{2} \ \rm g^{-1}$ \citep[][]{naknak94,semhenhel03}, Galactic disk surface density $\Sigma_{\rm gal}=1 M_{\odot} \rm pc^{-2}$ \citep{mckparhol15}, and $\phi_{\Sigma}{\sim}60$ (using a typical GMC density $\Sigma_{\rm cl}=60 M_{\odot} \rm pc^{-2}$; \cite{mivmurlee17}). These parameters applied to equation \ref{equation:eta_fir} imply that solar neighborhood clouds require $\eta_{\rm d}{\gtrsim}17$ to trap FIR.

Having mapped out the dust abundances and cloud bulk densities above which protoplanetary disks are shielded from external UV photons, but subject to external FIR heating, we next show that disk lifetimes can still be shortened in such fully optically thick ($\tau_{\rm UV}, \tau_{\rm cl, FIR}>1$) conditions. As displayed in Figure \ref{figure:tau}, external evaporation is shut down above $\eta_{\rm d}{\gtrsim}30\%$ in our fiducial GMC (and above ${\gtrsim}100\%$ and ${\gtrsim}10\%$ in our least and most dense clouds respectively), at which point $\tau_{\rm cl, FIR}{\sim}100$ (${\sim}300$ and ${\sim}30$ at each end of our GMC parameter space). In this regime, the background temperature of a protoplanetary disk will exceed the bolometric temperature of the cluster by a factor ${\sim}100^{1/4}{\sim}3$ (${\sim}500$ K for $T_{\rm bol}{\sim}150$ K). The disk viscosity will increase by the same factor, boosting the rate at which the disk viscously accretes. In addition to FIR heating, internal disk evaporation (e.g. through UV \cite[e.g.][]{gorhol09} or X-rays \cite[e.g.][]{picercowe19}) can still occur in the $\tau_{\rm UV}, \tau_{\rm cl, FIR}>1$ regime. We explore this scenario by setting $\Phi_{\rm FUV,\rm EUV}=0$ to evolve our disks only under bolometric heating and internal X-ray evaporation. We find a disk lifetime $\approx$0.8 Myr for a 500 K background temperature due mainly to increased stellar accretion, significantly shorter than the $\sim$several Myr lifetime of a cold ($\sim$30 K) disk. Determining the amplitude of this effect in the context of disk angular momentum transport via magnetic winds \citep{baisto13} would be a useful avenue for future work.

\subsection{Variations in Disk $\alpha$ Viscosity at Cosmic Noon}

%EJL
%Our protoplanetary disk evolution comparison between the primordial thick disk and solar neighborhood has thus far assumed 
We have thus far assumed
$\alpha=10^{-3}$ identically for both thick disk and the solar neighborhood.
%a constant Shakura-Sunyaev viscosity parameter \citep{shasun73} of $\alpha=10^{-3}$,
%EJL
%This choice of $\alpha$ is consistent with constraints from solar neighborhood disks \citep[e.g.][]{ros23}
While such a value is typically invoked for disks in the solar neighborhood (at least as an upper limit; see \citealt{ros23} for a review), it is not clear whether the same level of disk turbulence would manifest in the primordial thick disk.
%EJL
%This low value of $\alpha$ may reflect the fact that disks in the solar neighborhood are weakly ionized, and are thus unable to sustain accretion via the magneto-rotational instability \citep[MRI;][]{balhaw91} throughout the entirety of the disk \citep[since MRI requires ionized gas/magnetic field coupling; e.g.][]{gam96}. Solar neighborhood disks may be weakly ionized in part due to the strong reduction in cosmic ray flux produced by magnetized winds from host stars \citep[][]{cleadaber13}. Without cosmic rays, disks may suffer ionization through radionuclide decay \citep[e.g.][]{umenak09}, X-rays \citep[e.g.][]{glanaj97}, as well as stellar \citep[e.g.][]{perchi11} and interstellar \citep{cleadaber13} FUV photons. 
%EJL
For example, under the consideration of accretion mediated by magneto-rotational instability \citep[MRI;][]{balhaw91}, the intense ionizing radiation fields (cosmic ray and UV) present at cosmic noon could boost disk ionization, well beyond that in the solar neighborhood, and therefore increase MRI-generated $\alpha$.
%by facilitating MRI across a larger swath of the disk. 
%
In this section, 
%EJL
%our goal is therefore to 
%
we estimate over what spatial extent of the disk the MRI 
%EJL
%is liable to 
may
operate and thus potentially boost $\alpha$ above our fiducial value \cite[MRI in the ideal limit has been found to produce $\alpha{\sim}10^{-2}$, with a spread up to $\alpha{\sim}1$ depending on the strength and configuration of the magnetic field;][]{davstopes10,simbecarm12,lesfloerc23}. We will show that under the heightened cosmic ray flux in the primordial thick disk, the majority of the protoplanetary disk may be susceptible to MRI, potentially decreasing our fiducial disk lifetime to ${\lesssim}10^{5}$ yr.

In order for the MRI to operate, the instability's growth must outpace resistive damping. We follow the treatment of \cite{moherctur13} to roughly calculate the requisite ionization fraction for this to occur throughout the disk. 
%EJL
%Setting 
Comparing
the growth and resistive dissipation timescales 
%EJL
%equal
%
for the fastest growing mode yields the Elsasser number criterion \citep[equation 2 of][appropriate for local magnetic disturbances generated by MRI]{moherctur13}:

\begin{equation}\label{equation:elsasser}
    \Lambda_{\rm A}(R)\equiv\frac{v^{2}_{\rm A}(R)}{\eta(R) \Omega(R)}>1
\end{equation}
\noindent
where the Alfvén speed $v_{\rm A}(R)=B/\sqrt{4\pi \rho(R)}$, with $B$ the magnetic field, and where $\eta{\sim}230 x^{-1}_{\rm e} T^{1/2}$ cm$^{2}$ s$^{-1}$ is the Ohmic resitivity \citep{blabal94}, with $x_{\rm e}=n_{\rm e}/n_{\rm H}$ the electron number density fraction.\footnote{By using the total (mostly neutral) density $\rho$ in the Alfvén speed, we are implicitly assuming that ions exchange momentum with neutrals on a shorter timescale than the wave period \citep[${\sim}\Omega^{-1}$ for the fastest growing mode;][]{les21}. We confirm that the 
%EJL
%momentum transfer timescale of ions with neutrals 
ion-neutral collision timescale is indeed short
${\sim}(\gamma_{\rm i} \rho)^{-1}{\ll}\Omega^{-1}$, where $\gamma_{\rm i}=\big<\sigma v\big>/(\mu_{\rm i}+\mu_{\rm n})$, with the momentum exchange rate $\big<\sigma v\big>=1.6{\times}10^{-9}$ cm$^{3}$ s$^{-1}$ \citep[e.g.][]{warng99}, while $\mu_{\rm n}=2.33 m_{\rm H}$ and $\mu_{\rm i}=39 m_{\rm H}$ (assuming singly ionized potassium dominates the ion population, appropriate for thermally ionized regions \citep[e.g.][]{ume83,destur15}; at larger distances where thermal ionization is shut down, cosmic ray ionization of H$_{\rm 2}$ quickly produces ions like HCO$^{+}$ or Mg$^{+}$ via charge exchange \citep{destur15}, so that $\mu_{\rm i}{\sim}$20-30$m_{\rm H}$, which does not affect our conclusion).} The minimum ionization fraction for MRI to operate is then given by

\begin{equation}\label{equation:xe_mri}
    x_{\rm e, MRI}(R) \sim 115 \beta_{\rm pl}(R) \frac{\mu}{k_{\rm B}} \frac{\Omega(R)}{T^{1/2}(R)}
\end{equation}
\noindent
where the plasma $\beta_{\rm pl}$-parameter gives the ratio of thermal to magnetic pressure $\beta_{\rm pl}(R)=8\pi \rho(R) c^{2}_{\rm s}(R)/B^{2}$. 

\begin{figure}
\epsscale{1.2}
\plotone{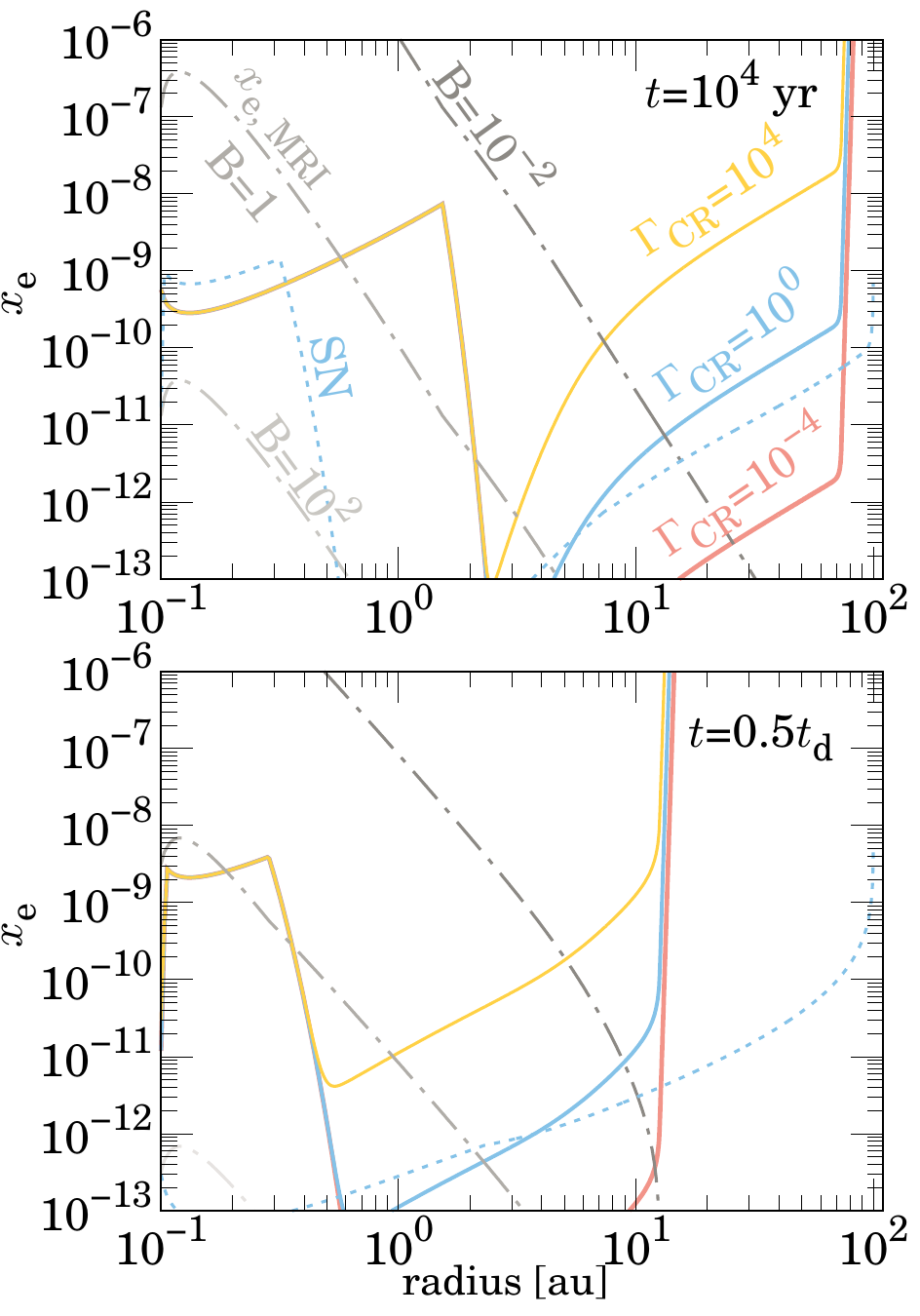}
\caption{Top: Midplane ionization fraction $x_{\rm e}=n_{\rm e}/n_{\rm H}$ versus radius for different cosmic ray exposures ($10^{4}, 10^{0}, 10^{-4} {\times}$ that of the local interstellar value in yellow, blue, and red curves respectively). Solid lines use the disk model from our fiducial thick disk simulation at time $t=1.2{\times}10^{4}$ yr, while the dashed ``SN" line uses our solar neighborhood disk at the same time. Cosmic ray ionization drops toward smaller radii due to attenuation through the disk 
%EJL
%(it grows 
(the ionization fraction shoots up
at the disk outer edge where the density $\rightarrow$0 due to evaporation). Electron fractions rise sharply inward of ${\sim}$2 au due to thermal ionization, before dropping again as density increases. The minimum ionization fraction needed to seed MRI in the face of resistive dissipation is plotted in grey dot-dashed lines, for different disk magnetic fields (assumed radially constant for each line; $B=10^{-2}, 1, 10^{2}$ G from dark to light grey respectively). Ionization levels must exceed the grey lines for MRI to operate; conditions at cosmic noon may facilitate MRI 
%EJL
%across much of the disk
even in the inner disk (inside 10 au)
if $B \geq 1$G and $\Gamma_{\rm CR} > 1$. 
Bottom: the same calculation, but at $t=0.5{\times}$ the disk lifetime $t_{\rm d}$ ($2.5{\times}10^{5}$ and $1.5{\times}10^{6}$ yr in thick disk and solar neighborhood). 
%EJL
The reduction of disk gas density allows for higher $x_{\rm e,CR}$ in the inner disk.
\label{figure:xe}}
\end{figure}

We will use equation \ref{equation:xe_mri} to chart out the minimum ionization fraction needed for MRI to operate at each location in the disk
%EJL
for a chosen $B\in [10^{-2},1,10^{2}]$ G.\footnote{Additional requirements for MRI include the criterion for generation of large-scale toroidal fields through Keplerian shear \citep[equation \ref{equation:elsasser} with $v^{2}_{\rm A}$ replaced with Keplerian speed $v^{2}_{\rm k}$;][]{tursan08}, and the ambipolar Elsasser number criterion \citep[equation \ref{equation:elsasser} with Ohmic resistivity $\eta$ replaced with the ambipolar resistivity $\eta_{\rm A}$;][]{war99,moherctur13}. The large-scale criterion is satisfied at lower ionization fractions than the Ohmic Elsasser number due to the longer dissipation time, so we do not compute it. We also confirm that wherever equation \ref{equation:elsasser} is satisfied, the ambipolar Elsasser number is almost always ${>}1$ (the exceptions occur for $\Gamma_{\rm CR}{\lesssim}10^{-1}$ in the outer disk).} 
%EJL
%In order to do so, we assume a range of magnetic fields $B\in [10^{-2},1,10^{2}]$ G. 
%
The lower limit is consistent with field strengths observed in local star-forming regions \citep[e.g.][]{girraomar06,belpadgir19,corsanhou21}, 
%EJL
and the fields within circumstellar disks can be
%since such fields will likely be 
%
amplified 
%EJL
from this background value
by shear and compression during disk formation \citep{war07}. Fields ${\sim}1$ G accord with constraints on the solar nebula's field \citep{weibaifu21}, while the upper limit follows from the requirement that $\beta_{\rm pl}{\lesssim}1$. We compute $x_{\rm e, MRI}$ across the disk for each fixed value of $B$ \citep[note that if angular momentum is transported via magnetic effects, $B$ likely increases toward smaller orbital radii;][]{war07}.

We next compare $x_{\rm e, MRI}$ to that produced by 
%EJL
cosmic ray and thermal
ionization $x_{\rm e, i}$. We compute the steady-state electron fraction due to cosmic ray ionizations at the disk midplane via $x_{\rm e,CR}(R){\sim}\sqrt{\zeta_{\rm CR}(R)/\alpha_{\rm r}n_{\rm H}(R)}$ \citep{froterbal02}, in which we have assumed pure molecular recombination at a rate $\alpha_{\rm r}=3{\times}10^{-6}T^{-1/2}$ cm$^{3}$ s$^{-1}$ \citep{oppdal74,spi78}. We set the cosmic ray ionization rate $\zeta_{\rm CR}(R)=\Gamma_{\rm CR}\zeta_{\rm CR,0}\exp{(-\Sigma(R)/2\lambda_{\rm CR})}$ (accounting for cosmic ray attenuation as in equation \ref{equation:cr}), with $n_{\rm H}(R)=\Sigma(R)/\sqrt{2\pi}H(R) m_{\rm H}$ the hydrogen number density. We also track thermal ionization following equation 20 of \cite{benkallug00}, relevant to viscously-heated innermost regions of temperature ${\gtrsim}10^{3}$ K. We sum the cosmic ray $x_{\rm e,CR}$ and thermal $x_{\rm e,T}$ ionization rates to get the total $x_{\rm e,i}=x_{\rm e,CR}+x_{\rm e,T}$.

We compare $x_{\rm e, MRI}$ to $x_{\rm e, i}$ in Figure \ref{figure:xe}. Assuming the magnetic field $B{\gtrsim}$1 G, disks at cosmic noon are sufficiently ionized to enable MRI across the majority of disk radii (${\gtrsim}$0.5 au), even if the cosmic ray flux is attenuated up to four orders of magnitude below the fiducial $\Gamma_{\rm CR}=10^{4}$.
%EJL
%(only $2{\lesssim}R{\lesssim4}$ au are too weakly ionized for MRI if $\Gamma_{\rm CR}=1$). 
Over time as the disk gas density depletes, the zone of high $x_e$ moves in, potentially enabling sustained MRI-driven turbulence (see bottom panel of Figure \ref{figure:xe}).
We explore how disk evolution will change in the case that MRI turbulence is facilitated across all disk radii by re-running our fiducial thick disk simulation with $\alpha=10^{-2}$. We find 
%EJL
further reduction in disk lifetime by factors of order unity (from ${\sim}5{\times}10^{5}$ yr to ${\sim}10^{5}$ yr),
%a shorter disk lifetime ${\sim}10^{5}$ yr, 
due to the increased stellar accretion rate (stellar accretion and external photoevaporation evacuate ${\sim}$half the disk mass each). Dedicated simulations tracking ionization and viscous evolution are needed to treat disks with $\alpha$ that varies radially (between weakly and strongly ionized regions) as well as temporally (as the disk cools, lowering $x_{\rm e, T}$, and density drops, increasing $x_{\rm e, CR}$).

\subsection{Giant Planet Formation}\label{sec:gas_giants}

Giant planet formation from core accretion should be inhibited in rapidly evaporating disks \citep[e.g.][]{winhawcol22}. If giant planets form in dense, evaporative stellar environments, an alternative channel may be via gravitational instability (GI). In order to collapse into a giant planet, a GI-unstable patch of the disk must radiate away the heat liberated from the heightened accretion/turbulence quickly enough to avoid thermal pressure halting further contraction; if this is not the case, the disk can self-regulate back to $Q\sim$1, maintaining a steady state of ``gravitoturbulence" \citep[][]{gam01}.

\cite{kramur11} suggest that externally irradiated disks may be \textit{more} susceptible to fragmentation than disks with cooling times set by internal dissipation. In the event that $Q\sim c_{\rm s}\Omega/G\Sigma$ is somehow driven ${<}1$, both types of disk experience increased pressure support from dissipation and turbulence that can counteract collapse. 
%EJL
%In regions of the disk where $T(R)$ is controlled by external heating rather than dissipation (i.e. $T(R){>}$the temperature of a purely viscously-heated region $T_{\rm vis}=(\frac{9}{8}\frac{\alpha k_{\rm B}}{\mu}\frac{\Sigma\Omega}{\sigma_{\rm SB}})^{1/3}{\sim}16$ K at 10 au; see equation \ref{equation:vis_dis}) the temperature is insensitive to an increase in dissipation rate: if GI-driven turbulence increases $\alpha$ by ${\sim}10{\times}$, the external temperature only needs to exceed $T_{\rm vis}$ by $10^{1/3}{\sim}2$ (${\sim}$30 K at 10 au) for irradiation to continue dominating the disk energy budget (our disks are irradiation-dominated for $R{\gtrsim}$a few au). 
%
Under dissipative heating, $T(R) \propto \alpha^{1/3}$ (see equation \ref{equation:vis_dis}) and so even if $\alpha$ increase by an order of magnitude, the disk temperature rises by just factors of $\sim$2. Such an increase can push $Q$ back up to $\geq$1 for disks dominated by viscous heating.
By contrast, in a region of the disk that was and remains dominated by external irradiation, $Q$ will remain below unity and GI can continue \citep[see also][]{kralod16}.
%
%Upon collapse, the disk temperature is largely unchanged and gravitational instability can continue. 
%
%In marginally unstable disks dominated by viscous heating, an increase in gas temperature by a factor ${\sim}$2 can push $Q$ back up to 1. To summarize: if $Q$ can be driven unstable in an irradiated disk, the instability cannot saturate \citep[see also][]{kralod16}.

At first blush, the typical background cluster temperatures we have found in this work $\sim$100 K are capable of driving $Q \gg $1 globally throughout the disk (for our initial disk masses $0.1 M_{\star}$), therefore shutting down any possibility of planet formation via fragmentation.  Planet formation via GI in such hostile environments must therefore proceed through an external agent driving $Q {<} 1$. One possibility is through intense mass infall onto the disk from the ambient envelope during the protostellar stage. In this case, fragment masses typically emerge 
%EJL
%$\gtrsim$ 
larger than
the deuterium burning limit, and can even continue growing as the disk rapidly accretes \citep[][]{kramuryou10}. It is unclear if early-onset GI from rapid infall should occur more or less frequently in the thick disk than in the thin disk, which depends on the ability of the Fe-poor, [$\alpha$/Fe]-enriched gas to cool (which in turn depends on the uncertain details of opacity).

Another possible mechanism is through tidal perturbations from stellar flybys. Theory and observations indicate that the tidal response of disks to stellar flybys can instigate accretion outbursts \citep[e.g.][]{pfatacste08,donliucue22}, and spiral arms \citep[e.g.][]{cuedipmen19,johhoabeu20}. Theoretical work has also shown that flybys can drive local enhancements in dust to gas ratio \citep[][]{cuedipmen19}, as well as direct gravitational fragmentation on large and small scales in the disk \citep[e.g.][]{thikrogoo10,mer15}.

Typical stellar densities in the nearest few pc surrounding a random star in highly correlated stellar environments are $\sim$10$^{4-5} M_{\odot}$ pc$^{-2}$ (still below the empirical maximum surface density observed in dense stellar systems $\sim$3$\times$10$^{5} M_{\odot}$ pc$^{-2}$; \cite{hopmurqua10}). The collision timescale reads \citep[][]{bintre08},

\begin{equation}\label{equation:t_col}
\begin{split}
    \hspace*{0cm} t_{\rm col} &\sim \bigg[4\sqrt{\pi} n_{\star}\sigma_{\star}d_{\rm enc}^{2}\bigg]^{-1} \\ 
    & \hspace*{-0.6cm} \sim 6 {\times} 10^{4} \bigg(\frac{n_{\star}}{10^{5} \ \mathrm{pc}^{-3}}\bigg)^{-1}\bigg(\frac{\sigma_{\star}}{1 \ \mathrm{km} \ \mathrm{s^{-1}}}\bigg)^{-1}\bigg(\frac{d_{\rm enc}}{10^{3} \ \mathrm{au}}\bigg)^{-2} \ \mathrm{yr},
\end{split}
\end{equation}

\noindent with $n_{\star}$ the stellar number density, $\sigma_{\star}$ the stellar velocity dispersion, and $d_{\rm enc}$ the encounter distance. Although dense stellar clusters born during cosmic noon could be short-lived due to rapid gas expulsion from massive star feedback \citep[e.g.][]{bankro13}, or
tidal perturbations by dense gas clumps \cite[e.g.][]{kru15} equation \ref{equation:t_col} indicates that a number of close flybys could still occur early on. To compare $t_{\rm col}$ to the tidal disruption timescale of stellar ``clusters" due to passing GMCs, we compute $t_{\rm dis}{\sim}|E_{\rm cl}/\dot{E}_{\rm cl}|$, with the cluster energy per unit mass $|E_{\rm cl}|= G M_{\star, \rm cl}/r_{\rm cl}$ where $M_{\star, \rm cl}=\epsilon_{\star}M_{\rm cl}$ is the cluster stellar mass. The energy injection rate per unit mass due to perturbations is (\cite{gieporbau06}; see also \cite{bintre08}),

\begin{equation}\label{equation:edot}
\begin{split}
    \dot{E}_{\rm cl}&=f\frac{G^{2}\rho_{\rm gal}\Sigma_{\rm cl}\big<r^{2}_{\star}\big>}{\sigma_{\rm gal}} \\ 
\end{split}
\end{equation}
\noindent 
where the prefactor $f=\frac{4}{3}\pi^{3/2}g\phi_{\rm ad}$, with $g=1.5$ a correction for the spatial extent of the perturbing GMC \citep{kru12}, and $\phi_{\rm ad}$ corrects for energy absorption through adiabatic expansion of the cluster if the perturbation is slower than the cluster's crossing time \citep{wei94}. Following \cite{kru15}, we thus take $\phi_{\rm ad}=(1+(t_{\rm pert}/t_{\rm c})^{2})^{-3/2}$, where $t_{\rm pert}=(4\sqrt{\pi}n_{\rm GMC}\sigma_{\rm gal}d^{2}_{\rm pert})^{-1}$ is the perturbation duration (assumed to be of order the time between GMC encounters), with $n_{\rm GMC}=\rho_{\rm gal}/M_{\rm cl}$ the number density of GMCs in the Galactic disk, where $\rho_{\rm gal}{\sim}\Sigma_{\rm gal}/\sqrt{2\pi}H_{\rm gal}{\sim}G\Sigma^{2}_{\rm gal}/\sqrt{2\pi}\sigma^{2}_{\rm gal}$ is the bulk ISM density in the Galactic disk and $H_{\rm gal}=\sigma^{2}_{\rm gal}/G\Sigma_{\rm gal}$ the disk scale height, $d_{\rm pert}=n^{-1/3}_{\rm GMC}$ the typical inter-GMC distance, and $t_{\rm c}=r_{\rm cl}/\sigma_{\rm cl}$ the cluster crossing time. We set the mass-weighted mean-square radius of stars in the cluster $\big<r^{2}_{\star}\big >{\sim}\frac{1}{3} r^{2}_{\rm cl}$, and $\sigma_{\rm gal}$ is assumed to be the velocity dispersion between GMCs and our cluster. With this formulation, $t_{\rm dis}$ reads,

\begin{equation}\label{equation:t_dis}
\begin{split}
    t_{\rm dis}&=\frac{1}{f}\frac{M_{\star,\rm cl}}{\Sigma_{\rm cl}\big<r_{\star}^{2}\big>}\frac{\sigma_{\rm gal}}{G\rho_{\rm gal}r_{\rm cl}} \\ 
    & \hspace{-0.5cm} {\sim}3.3 \bigg(\frac{\epsilon_{\star}}{0.1}\bigg) \bigg(\frac{\sigma_{\rm gal}}{45 \ \rm{km \ s^{-1}}}\bigg)\bigg(\frac{\Sigma_{\rm gal}}{1.7{\times}10^{3} \ M_{\odot} \ \rm pc^{-2}}\bigg)^{-1}\phi^{-1}_{\rm ad} \  \mathrm{Myr}\\
\end{split}
\end{equation}
\noindent 
where we have rewritten $r_{\rm cl}=(M_{\rm cl}/\pi \Sigma_{\rm cl})^{1/2}$, and $M_{\rm cl}=M_{\rm J}$ via equation \ref{equation:jeans}. These choices produce $\phi_{\rm ad}{\sim}0.85$ and $t_{\rm dis}{\sim}4$ Myr. We note that, given the ${\propto}r^{-3}_{\rm cl}$ dependence of equation \ref{equation:t_dis}, self-consistent dynamical calculations tracking the cluster radius would be needed to estimate $t_{\rm dis}$ to better than an order of magnitude. Equation \ref{equation:t_dis} indicates that $t_{\rm dis}/t_{\rm col}{\sim}10^{2}$, so ${\sim}$one hundred close encounters between a subject star and nearby stars could occur before the ``cluster" is tidally disrupted. On the other hand, equation \ref{equation:t_col} yields $t_{\rm d}/t_{\rm col}{\sim}10$ (where $t_{\rm d}$ is the protoplanetary disk lifetime), so ${\sim}$ten GI-triggering encounters could occur before the disk evaporates. 

Giant planet formation from GI has not been studied in tandem with external evaporation. The final mass of an accreting body formed through GI will be strongly limited as the disk rapidly evaporates \citep[as also suggested in the context of internal evaporation by][]{kramuryou10}. Given that disk radii and masses are typically truncated to $\sim$10$\%$ of their initial values during the first few $\sim$10$^{5}$ yr in the thick disk (see e.g. Figure \ref{figure:mass_compare}), we expect any planets born through this flavor of GI to be a) relatively close in ($\lesssim$20 au), and b) possibly less massive than the typical $\sim$brown dwarfs born through traditional GI, owing to the reduced fragmentation mass in disks of small radius outweighing the increase in disk temperature ($M_{\rm frag}{\sim}\Sigma H^{2}\propto \Sigma T R^{3} \propto R^{3/2}$ assuming $T \propto R^{-\beta}$, $\Sigma\propto R^{\beta-3/2}$ as in equation \ref{equation:IC}). 

\subsection{Observational Predictions}\label{sec:predictions}

We now outline how our results may bear on the planet population around thick disk stars. We summarize possible avenues below, connecting to current or near future observational surveys where possible.  

\subsubsection{The Abundance of Gas-Rich Planets}\label{sec:extended_drop}

With a shorter disk lifetime available for accretion of massive gaseous atmospheres, sub-Saturn and gas giant occurrence rates should suffer a greater reduction from thin to thick disks compared to small planets. Although external disk evaporation proceeds outside-in, our results highlight that disks continue to erode under evaporation and accretion even once the disk is ${\lesssim}$a few au in size. Both long and short period (${\lesssim}$300 days) gaseous planets are therefore likely to be affected by disks that only live ${\sim}$a few${\times}10^{5}$ yr. The outside-in evaporation of protoplanetary disk gas implies that the drop in gas-rich planet occurrence compared to small planets 
%EJL
%should extend and likely grow beyond periods 
should be particularly severe at periods
$\gtrsim$300 days. 
%EJL
Unlike the solar neighborhood where we observe
%The measured 
a peak in giant planet occurrence at ${\sim}10^{3}$ days \citep[e.g.][]{fermulpas19,niedermac19,fulroshir21}, 
%EJL
thick disk giants may not feature such a sharp peak
%may exhibit a significant reduction 
%
due to preferential destruction of disk outer radii under evaporation. 

Lower abundance of gas-rich planets around thick disk stars may be degenerate with the well-established giant planet \citep{fisval05} and sub-Saturn \citep[][]{petmarwin18} occurrence-metallicity ([Fe/H]) correlations, which have been interpreted as a result of fewer planetary building blocks in low-metallicity disks \citep[e.g.][]{idalin04b}. 
Evaporation and a lack of building blocks may be disentangled by isolating the effect of [$\alpha$/Fe] on gas-rich planet occurrence: since elevated [$\alpha$/Fe] in Milky Way disk stars indicates rapid star formation, as implied by the spatial uniformity and high value of the [Fe/H] at which the ``knee" in [$\alpha$/Fe] occurs (\cite{nidbovbir14}; see also \cite{gilwys98}), primordial disk evaporation predicts that gas-rich planet occurrence should anti-correlate with [$\alpha$/Fe], at a given [Fe/H] (for the range of metallicities exhibited in the thick disk, $-1{\lesssim}$[Fe/H]${\lesssim}+$0.2; e.g. \cite{nidbovbir14,leubovmac23}).
%EJL
A %possible caveat 
complication
to this prediction however is that common molecules involving $\alpha$ elements C and O are efficient coolants in photoevaporating gas \citep{glanajige04,erccla10}, which could decrease photoevaporative mass loss rates. Future work should address to what degree photoevaporative mass loss rates decline due to an increase in [$\alpha$/Fe]. 

The imprint of disk evaporation may also be discerned by searching for an increase in the ratio of rocky to gas-rich planet occurrence
%EJL
around thick disk stars.
If massive (${\gtrsim}10 M_{\oplus}$) rocky cores can assemble around thick disk stars (super-Earth/sub-Neptune occurrence is not zero, so rocky cores \textit{do} form), they should remain gas-poor. Such failed giant planets will likely still retain modest gas envelopes, which is a robust outcome even in extremely gas-poor disks \citep{leechiorm14,leechi15,leechi16}. 

%EJL
%Since both distant and close-in gaseous planets are likely affected by disk evaporation, observational surveys probing long and short periods can each help constrain our theory. The occurrence of short period gas-rich planets around thick disk stars can potentially be measured by the \textit{Transiting Exoplanet Survey Satellite (TESS)}, which is actively expanding the census of thick disk planetary systems \citep[e.g.][]{ganshpliv20,briweidai23}. The \textit{Nancy Grace Roman Space Telescope (Roman)} is also projected to discover ${\sim}10^{4}$ close-in, gas-rich planets through its transit survey \citep{monyeepen17,wilbarpow23}.
The census of small and large planets at short orbital periods around thick disk stars can be expanded with current and upcoming transit survey including the \textit{Transiting Exoplanet Survey Satellite (TESS)} \citep[e.g.][]{ganshpliv20,briweidai23}, \textit{PLATO} \citep{rauaercab24}, and the \textit{Nancy Grace Roman Space Telescope (Roman)} \citep{monyeepen17,wilbarpow23}.
%The sheer number statistics afforded by a survey of this magnitude will provide broad coverage of thick and thin disk FGK stars \citep{wilbarpow23}. 
At orbital periods beyond $\sim$300 days, \textit{Roman}'s microlensing survey will be crucial in determining the relative population of rocky and giant planets between thin and thick disk. In terms of constraining the thin vs.~thick disk population of wide-orbit giants, \textit{Gaia} Data Release 4 may prove handy \citep{perharbak14}.
%\textit{Roman}'s microlensing survey may also constrain the occurrence of gas-rich planets on wide orbits (${\gtrsim}$1 au) around thick disks stars. Similarly, \textit{Gaia} Data Release 4 may contain a census of astrometrically-detected long-period giant planets around thick disk stars \citep{perharbak14}, which could help constrain their abundance. The \textit{PLATO} mission may complement \textit{Roman} and/or \textit{Gaia} through its transit survey, which will be capable of detecting gas giants with orbital periods ${\lesssim}$a few years \citep{rauaercab24}.  Gas-rich planet occurrence data from \textit{TESS}, \textit{Roman}, \textit{Gaia}, and/or \textit{PLATO} can be connected to our theory by comparing to the predicted occurrence from [Fe/H] alone, or by searching for a trend with [$\alpha$/Fe] (assuming all host targets have well-measured metallicities).

Signatures of inhibited gas-rich planet formation around thick disk stars may already be present in the data: \cite{swabannar23} report that stars hosting giant planets (from hot to cold Jupiters at ${\sim}$a few au) appear to preferentially exhibit thin disk compositions, kinematics, and ages (from spectroscopic, kinematic, and isochrone measurements derived from \textit{Gaia} DR3). \cite{chexiezho23} also find that hot Jupiters are more common around thin disk stars than thick disk \cite[identified using stellar kinematics;][]{chexiezou21}, although they do not report a significant trend with stellar kinematic age for warm and cold Jupiter occurrence.

\subsubsection{\texorpdfstring{$\bar{n}_{\rm p}$}{Planets per Star} vs. \texorpdfstring{$F_{\rm p}$}{Stars with Planets}}

\cite{baszuc22} report that thick disk planet occurrence, integrated over all planet types from 0.5--10$R_{\oplus}$, decreases in both $\bar{n}_{\rm p}$ and $F_{\rm p}$. The occurrence decreases less in $F_{\rm p}$, however. Using independent occurrence rate methodology, \cite{yanchexie23} on the other hand do not report a significant drop in $F_{\rm p}$ with stellar kinematic age. We will next discuss how differences in measured $\bar{n}_{\rm p}$ vs. $F_{\rm p}$ could constrain the mechanism reducing thick disk planet occurrence.

Assuming that $\bar{n}_{\rm p}$ decreases by a larger amount than $F_{\rm p}$, as found by \cite{baszuc22}, implies that the average planet multiplicity, $\bar{m}_{\rm p}$ (=$\bar{n}_{\rm p}/F_{\rm p}$), also decreases. The \cite{baszuc22} data therefore imply that the underlying physical mechanism(s) must therefore be capable of suppressing the formation of planetary systems entirely (since the number of planetary systems $\propto F_{\rm p}$), as well as inhibiting multiplicity. External evaporation may be capable of preventing the formation of planetary systems entirely (decreasing $F_{\rm p}$), in the shortest lived disks with $t_{\rm d} {\sim}10^{5}$ yr. Thus, the measurement by \cite{baszuc22} of a decrease in $F_{\rm p}$ may be qualitatively consistent with disk evaporation. On the other hand, if the dearth of planets is related to dynamical instability, we may to first order expect the total number of planetary systems $\propto {F}_{\rm p}$ to remain the same (unless the instability is wholly destructive), but the multiplicity to be reduced by the same amount as $\bar{n}_{\rm p}$. The measurement by \cite{yanchexie23} that $F_{\rm p}$ is insensitive to stellar age may therefore be more consistent with dynamical instability rather than early protoplanetary disk destruction.

A potential complication to this scenario however is that disk evaporation itself can incite dynamical instabilities that could affect multiplicity \cite[e.g.][] {war81,petmunkra20}. In order to pry apart dynamical evolution and disk evaporation, more theoretical work is therefore needed to understand how \textit{external} evaporation affects post-disk multiplicities and architectures (i.e. a low-mass disk truncated down to ${\sim}$au scales, vs. a more massive transition disk with an inner hole, as envisaged by e.g. \cite{petmunkra20}). Further theoretical work tying external disk evaporation to the planet formation process is also needed to determine the precise expected outcome in $\bar{m}_{\rm p}$ vs. $F_{\rm p}$. 

%EJL
%More precise observational measurements of $F_{\rm p}$ and $\bar{m}_{\rm p}$ for each individual planet class would be useful to make a conclusive comparison to theory (\cite{baszuc22} cannot distinguish differences in $F_{\rm p}$ between individual planet classes).

\subsubsection{Planetary Atmosphere Compositions}\label{sec:atmospheres}

The compositions of planetary atmospheres may also differ if they were accreted in strongly heated protoplanetary disks. Under standard disk conditions, the carbon to oxygen ratio (C/O) of disk gas is thought to increase away from the star at early times as the temperature drops and major oxygen-carrying species freeze out (primarily water ice at $\sim$150 K, and CO$_{2}$ at $\sim$50 K; \cite{obemurber11}). As the disk cools, the C/O ratio can exhibit non-monotonic behavior across the disk however, obfuscating the relation between planetary C/O and formation location \citep{molmolbit22}.

Without major ice lines (see Figure \ref{figure:Tbolt} and surrounding discussion), the C/O ratio planets can accrue from disk gas may be roughly constant throughout the disk, and the abundance of gaseous C and O could remain high everywhere. We may therefore expect C/O ratios in the atmospheres of planets around thick disk stars to be closer to the stellar value (\textit{not} the solar value - thick disk stars have [C/O]${\sim}-$0.2, varying with [Fe/H]; \cite{framordim21}), and to possibly exhibit more uniformity, than those around thin disk stars. Relative trends laid down at birth between thin/thick disk planetary atmospheres could be obscured by long-term chemical processes however. Differences in initial condition could be erased over ${\sim}$10 Gyr due to, e.g., atmospheric pollution \citep[][]{kurhorogi23}, escape-driven fractionation \citep{huseayun15}, exchange with the planetary interior \citep[][]{chafalsos23}, or UV-driven chemical reactions \citep{madagumos16}. Even during planet formation, chemical evolution of the disk in space and time (e.g. cooling due to cluster expansion) could drive the C/O ratio away from the stellar value \citep[e.g.][]{eiswalvan18}. How each of these processes affect relative trends between thin and thick disk planets should be addressed in future work.

Planetary atmospheres in the thick disk are observable by the \textit{James Webb Space Telescope} ({\it JWST}; \cite{grelinmon16}). As a first step to determine how many thick disk planets orbit stars bright enough to allow {\it JWST} characterization, we have compiled a list of thick disk planets (or thick disk candidates) from the literature with strong Transmission Spectroscopy Metrics (TSM) \citep[as defined by ][]{kembealou18}. The TSM is a simple, empirical measure of the signal to noise expected for a transmission measurement of a given planet observed by \textit{TESS}. A TSM $>$92 for planet radii 1.5$<R_{\rm p}<$2.75$R_{\oplus}$, or TSM $>84$ for 2.75$<R_{\rm p}<$4$R_{\oplus}$, indicate a strong candidate \citep{kembealou18}. Our list contains TOI 4438 b (TSM${\sim}$135; \cite{gofchamur24}), TOI 2406 b (TSM${\sim}$115; \cite{welracsch21}), TOI 561 c (TSM${\sim}$110; \cite{lacwilmal22}), TOI 2018 b (TSM${\sim}$103; \cite{daischreg23}), and possibly TOI 1730.01 (TSM${\sim}150$, adopting the \textit{TESS} data from the
\hyperlink{https://exoplanetarchive.ipac.caltech.edu/overview/TOI-1730}{Exoplanet Archive}).\footnote{Note that TOI 561 b is already slated for JWST characterization - see \url{https://www.stsci.edu/jwst/science-execution/approved-programs/general-observers/cycle-2-go}.}

Of the planets in our list, only TOI 561 and TOI 2018 have host stellar radii that overlap with the $K2$ stellar sample from the occurrence measurement of \cite{zinharchr23} (0.843$\pm$0.005$R_{\odot}$, and $0.62{\pm}0.01R_{\odot}$, respectively). The remaining targets orbit M stars. The small overlap between atmospheric targets and the $K2$/\textit{Kepler} sample may prevent a one-to-one comparison to the theoretical calculations we have presented, which focus on Solar-type stars. We caution that our theoretical results may not apply to thick disk M stars, due to differences in mass segregation between Solar-type and lower mass stars in clusters. M-type stars are observed to harbor larger disk fractions than Solar-type stars (in nearby clusters \citep{pfadehmic22}, as well as denser environments \citep{fanboekin12}), an observation that is at odds with photoevaporation theory, which predicts stronger mass loss around lower mass stars \citep[e.g.][]{hawclarah18}. Mass segregation of M stars towards the outskirts of clusters \citep[e.g.][]{pfastemen14} could reconcile photoevaporation theory with these observations, which our theory has not captured. Ongoing \textit{TESS} discoveries of thick disk planets may find more G and K-type targets to allow more direct comparison to our theory. 

\subsection{Forming Small Planets}

The model presented in this paper is intended to serve as a first outline of the relevant physical processes differentiating protoplanetary disk evolution in the primordial thick disk from the solar neighborhood. This paper has therefore not explicitly addressed how differences in protoplanetary disk evolution pathways may inhibit super-Earth/sub-Neptune formation. Below we outline the possible leading-order mechanisms inhibiting super-Earth/sub-Neptune formation.

The formation of super-Earths/sub-Neptunes differs from that of giant planets in that a smaller reservoir of disk material is required to create them (in the case of super-Earths, only ${\sim}$a few $M_{\oplus}$ of dust is called for). Photoevaporation can affect the dust budget in a disk by directly advecting grains in the flow out to space \citep[e.g.][]{selboocla20}, or by stranding dust at large distances where gas has been evaporated \citep{qiacolhaw23}. Both processes can only affect dust grains that have not grown sufficiently large to resist entrainment in the wind \citep{facclabis16}, or rapidly migrate inwards under aerodynamic coupling to the gas \citep{hawclarah18}. The timescale over which grains in the outer disk ($\sim$50 au) remain small enough to be affected by evaporation is ${\sim} 5 {\times} 10^{4}$ yr (assuming a grain fragmentation velocity of $\gtrsim$10 m s$^{-1}$ \citep{birklaerc12}, which yields a dust loss timescale in the outer disk that is set only by the time to coagulate above the maximum entrainment size; e.g. \cite{selboocla20}). The protoplanetary disk evolution illustrated in Figures \ref{figure:mass_compare} and \ref{figure:rdisk_tdisk} respectively highlights that even by this time, ${\sim} 60 \%$ of the gas disk mass is lost, and the disk radius is truncated to ${\sim}$10 au (assuming a lower fragmentation velocity ${\sim}1$ m s$^{-1}$ stunts grain growth below the maximum entrainment size beyond ${\sim}$30 au while also decreasing radial drift, lengthening the dust loss timescale to the timescale for gas disk evaporation ${\sim}10^{5}$ yr). We therefore expect a significant fraction of the dust initially beyond ${\sim}$10 au (${\sim}200 M_{\oplus}$) to be affected by the wind before it is able to migrate inward and potentially form planets. We plan to address the evolution of dust in a follow-up paper.

\section{Conclusion}\label{sec:conclusion}

Recent exoplanet demographics studies have revealed that super-Earths and sub-Neptunes exhibit an unexpected stellar type dependence: they are $\sim$50-70$\%$ less common around stars in the Galaxy's thick disk than those in the thin disk \citep[][]{baszuc22,zinharchr23}. In this work we have explored how protoplanetary disk evolution may have differed around stars born in the thin vs.~primordial thick disk. We summarize the two key results of this investigation below:

\begin{itemize}
    \item The primordial thick disk was hostile to planet formation. Protoplanetary disks at cosmic noon were subjected to radiation fields up to $\sim$7 orders of magnitude more intense than in present-day solar neighborhood GMCs. Protoplanetary disks photoevaporate to destruction in ${\sim}$0.5 Myr, limiting the time allowed for the onset of planet formation. 
    %EJL
    %We therefore propose that the planet occurrence deficit discovered by \cite{baszuc22} and \cite{zinharchr23} may be primordial in origin.
    %
    \item Protoplanetary disks at cosmic noon were heated to temperatures above major volatile species' sublimation temperatures. All thick disk conditions we consider sublimate CO, CH$_{4}$, and N$_{2}$ throughout the entire disk. In the most extreme environments, even water is vaporized for $\gtrsim$Myr.
\end{itemize}

Our theory makes several observational predictions that can elucidate how protoplanetary disk evaporation may sculpt planet demographics around thick disk stars. These are testable by current or near-future exoplanet surveys:

\begin{itemize}
    \item Gas-rich planets should suffer a larger decrease in occurrence than super-Earths/sub-Neptunes, 
    %EJL
    %and this feature should extend or grow beyond $\gtrsim$300 days; the occurrence ratio of super-Earths and sub-Neptunes to gas-rich planets should grow around thick disk stars.
    so that the ratio of small to large planets is enhanced around thick disk stars, particularly at long orbital periods.
    \item The occurrence deficit should also be evident in fraction of stars with planets $F_{\rm p}$. 
    %EJL
    %Decreases in occurrence should be evident in each planet type individually.
    \item Atmospheric C/O ratios may be closer to the stellar value ([C/O]${\sim}-$0.2), and possibly exhibit more uniformity, in thick disk planets relative to thin disk. 
\end{itemize}

The results of this paper suggest that protoplanetary disk evolution -- and by extension, planet formation -- may be intimately connected to Galactic evolution via the star formation process. Future observational surveys will elucidate how exoplanetary demographics and planet formation have evolved over cosmic time.

\acknowledgments

We thank the referee for a prompt and insightful report that improved the paper. T.H. would like to thank Dolev Bashi, Jo Bovy, Marta Bryan, Yayaati Chachan, Christine Mazzola Daher, Thomas Haworth, Kaitlin Kratter, Diederik Kruijssen, James Owen, Johanna Teske, Jiwei Xie, and Claire Ye for insightful conversations and correspondence. E.J.L. thanks Philip Hopkins, Nadine Soliman, and Jon Zink for useful discussions. T.H. acknowledges support from an NSERC Alexander Graham Bell CGS-D scholarship. E.J.L. gratefully acknowledges support by NSERC, by FRQNT, by the Trottier Space Institute, and by the William Dawson Scholarship from McGill University. Figures were produced using \texttt{gnuplot}. This research was enabled in part by support provided by Calcul Québec (\url{https://www.calculquebec.ca/}), Compute Ontario (\url{https://www.computeontario.ca/}), the BC DRI Group, and the Digital Research Alliance of Canada (\url{https://alliancecan.ca/}).

\bibliography{hallatt}{}
\bibliographystyle{aasjournal}

\section{Appendix A: Disk Evolution}\label{sec:appendix}

\begin{figure*}\label{mdot_efficiency}
\centering
\includegraphics[width=1.05\textwidth]{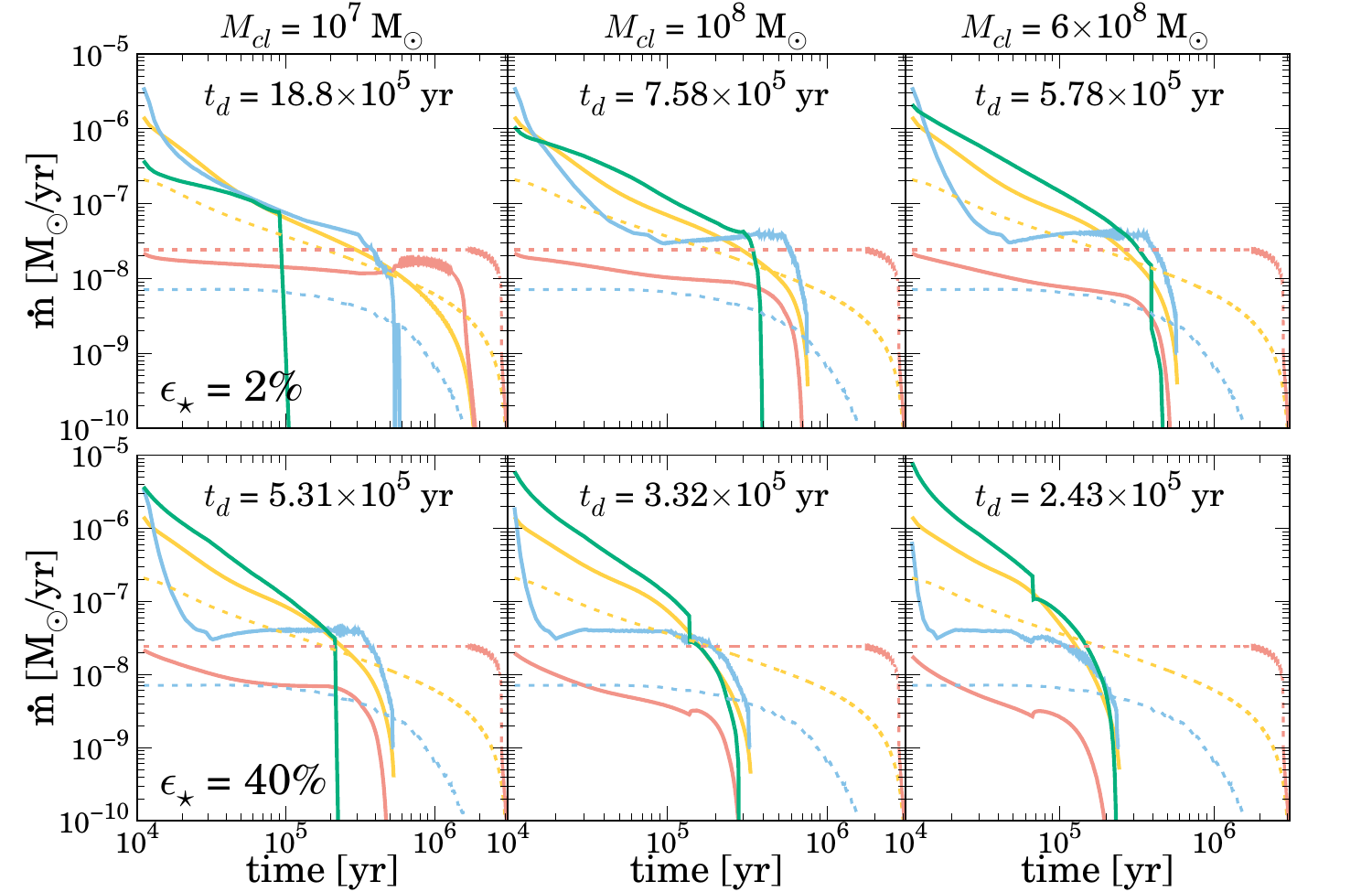}
\caption{Mass loss rate evolution for disks in each cloud, with varying star formation efficiency $\epsilon_{\star}=2\%$ (top row) and $\epsilon_{\star}=40\%$ (bottom row). Colors correspond to mass loss channels as in Figure \ref{figure:mdcomp}. The
solar neighborhood evolution is plotted with dashed lines. 
\label{figure:mdot_paramspace_eff}}
\end{figure*}

\begin{figure*}\label{mdot_segregation}
\centering
\includegraphics[width=1.05\textwidth]{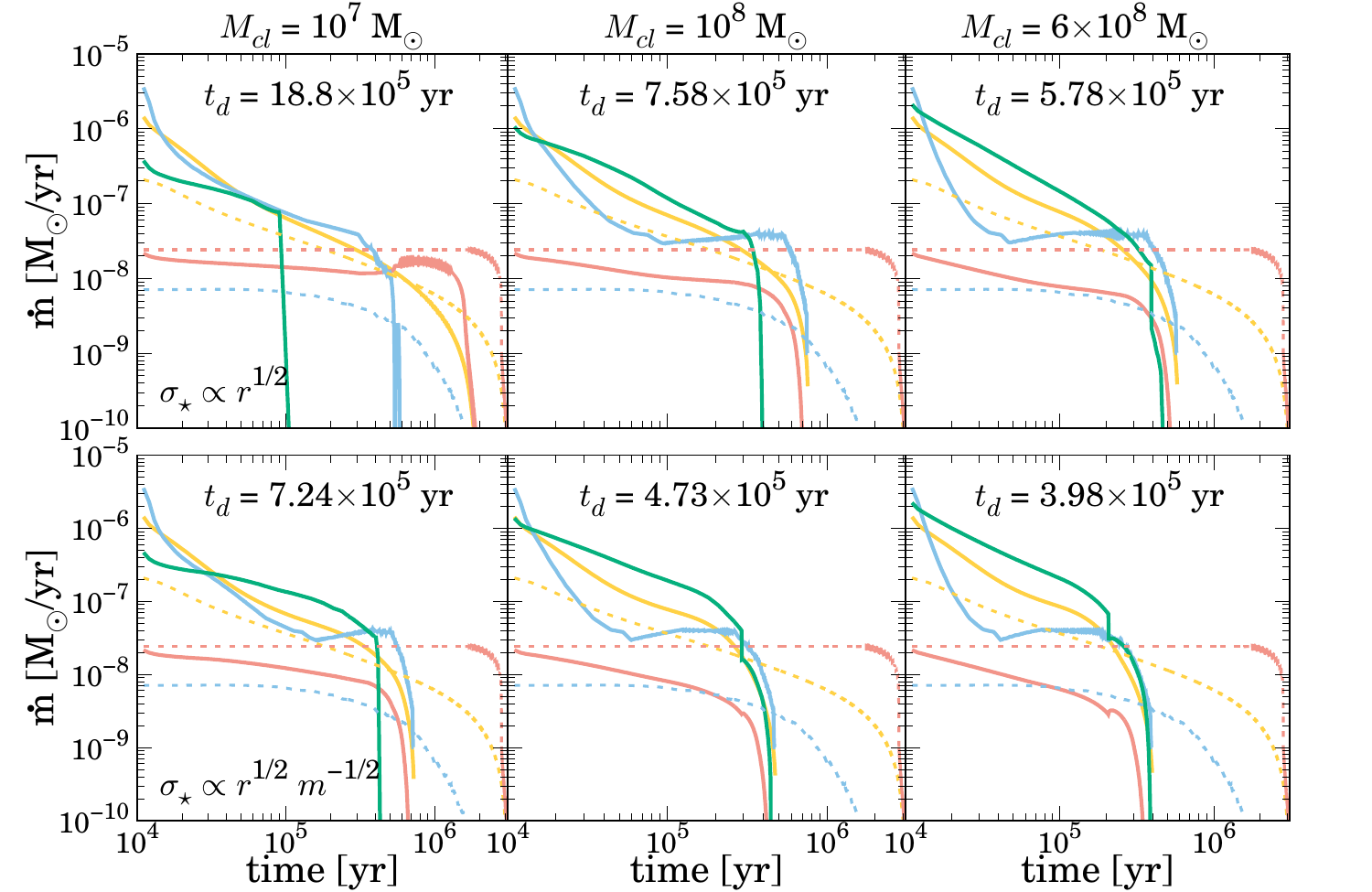}
\caption{Mass loss rate evolution for disks in each cloud, with varying mass segregation parameter $\beta=0$ (top row) and $\beta=1/2$ (bottom row). 
\label{figure:mdot_paramspace_seg}}
\end{figure*}

In this appendix we display supplemental figures displaying disk mass loss rate evolution across our parameter space. Figure \ref{figure:mdot_paramspace_eff} charts evolution in environments with low and high star formation efficiency $\epsilon_{\star}=2\%,\ 40\%$. Figure \ref{figure:mdot_paramspace_seg} shows the effect of varying the level of mass segregation for clouds of each mass, at fixed star formation efficiency.

\end{document}